\documentclass[%
reprint,
superscriptaddress,
amsmath,
amssymb,
aps,
pra,
longbibliography,
floatfix,
]{revtex4-2}

\usepackage{times}
\usepackage[colorlinks=true,linkcolor=blue,urlcolor=blue,citecolor=blue]{hyperref}
\usepackage{graphicx}
\usepackage{bm}
\usepackage{bbold}
\usepackage{dsfont}
\usepackage{xcolor}
\usepackage[normalem]{ulem}

\usepackage{scalerel}
\usepackage{tikz}
\usetikzlibrary{svg.path}
\definecolor{orcidlogocol}{HTML}{A6CE39}
\tikzset{
orcidlogo/.pic={
	\fill[orcidlogocol] svg{M256,128c0,70.7-57.3,128-128,128C57.3,256,0,198.7,0,128C0,57.3,57.3,0,128,0C198.7,0,256,57.3,256,128z};
	\fill[white] svg{M86.3,186.2H70.9V79.1h15.4v48.4V186.2z}
	svg{M108.9,79.1h41.6c39.6,0,57,28.3,57,53.6c0,27.5-21.5,53.6-56.8,53.6h-41.8V79.1z M124.3,172.4h24.5c34.9,0,42.9-26.5,42.9-39.7c0-21.5-13.7-39.7-43.7-39.7h-23.7V172.4z}
	svg{M88.7,56.8c0,5.5-4.5,10.1-10.1,10.1c-5.6,0-10.1-4.6-10.1-10.1c0-5.6,4.5-10.1,10.1-10.1C84.2,46.7,88.7,51.3,88.7,56.8z};
}
}
\newcommand\orcid[1]{\href{https://orcid.org/#1}{\mbox{\scalerel*{
			\begin{tikzpicture}[yscale=-1,transform shape]
				\pic{orcidlogo};
			\end{tikzpicture}
		}{R}}}}

\begin{document}
\title{Accelerated quantum control in a three-level system by jumping along the geodesics}

\author{Musang Gong\orcid{0000-0002-3357-5543}}
\author{Min Yu}
\author{Ralf Betzholz\orcid{0000-0003-2570-7267}}
\author{Yaoming Chu\orcid{0000-0003-2881-0040}}
\author{Pengcheng Yang}
\email{pcyang@hust.edu.cn}
\affiliation{School of Physics, Hubei Key Laboratory of Gravitation and Quantum Physics, Institute for Quantum Science and Engineering, Huazhong University of Science and Technology, Wuhan 430074, China}
\affiliation{International Joint Laboratory on Quantum Sensing and Quantum Metrology, Huazhong University of Science and Technology, Wuhan 430074, China}
\author{Zhenyu Wang}
\affiliation{Guangdong Provincial Key Laboratory of Quantum Engineering and Quantum
Materials, School of Physics and Telecommunication Engineering, South
China Normal University, Guangzhou 510006, China}
\affiliation{Frontier Research Institute for Physics, South China Normal University,
Guangzhou 510006, China}
\author{Jianming Cai}
\affiliation{School of Physics, Hubei Key Laboratory of Gravitation and Quantum Physics, Institute for Quantum Science and Engineering, Huazhong University of Science and Technology, Wuhan 430074, China}
\affiliation{International Joint Laboratory on Quantum Sensing and Quantum Metrology, Huazhong University of Science and Technology, Wuhan 430074, China}
\affiliation{Shanghai Key Laboratory of Magnetic Resonance, East China Normal University, Shanghai 200062, China}

\begin{abstract}
In a solid-state spin system, we experimentally demonstrate a protocol for quantum-state population transfer with an improved efficiency compared to traditional stimulated Raman adiabatic passage (STIRAP). Using the ground state triplet of the nitrogen-vacancy center in diamond, we show that the required evolution time for high-fidelity state transfer can be reduced by almost one order of magnitude. Furthermore, we establish an improved robustness against frequency detuning caused by magnetic noise as compared to STIRAP. These results provide a powerful tool for coherent spin manipulation in the context of quantum sensing and quantum computation.
\end{abstract}

\maketitle
\date{\today}

\section{Introduction}
Coherent control and manipulation with a high fidelity of the quantum state of a system has long been the subject of intensive research in modern quantum technologies, such as coherent manipulation of atomic and molecular systems~\cite{kral2007colloquium,bassett2014ultrafast}, quantum information processing~\cite{farhi2001quantum,monroe2013scaling,chen2020parallel}, and high-precision measurement~\cite{kasevich2002coherence,kotru2015large,buckley2010spin,liu2020super}. There is a vast literature proposing and implementing various methods for this purpose, such as adiabatic-passage techniques~\cite{shapiro2000coherent,rangelov2005stark,torosov2011high,kovachy2012adiabatic}, which are robust against variations of the control fields~\cite{bergmann1998coherent,fubini2007robustness,kumar2016stimulated,du2016experimental,kandel2021adiabatic} and have been widely applied, in quantum-state engineering~\cite{vitanov2001laser}, quantum simulation~\cite{aspuru2005simulated,kim2010quantum,biamonte2011adiabatic}, and quantum computation~\cite{jones2000geometric,barends2016digitized}, to mention only a few.

Among these adiabatic-passage techniques, stimulated Raman adiabatic passage (STIRAP)~\cite{gaubatz1990population,vitanov2017stimulated} is a paradigm example for adiabatic population transfer between two distinct states in a three-state system without populating the intermediate state using two control fields. Due to the robustness against control-parameter perturbations and the relaxation through spontaneous emission of the intermediate state, STIRAP has been extensively used in the realization of various quantum-information-processing tasks~\cite{kis2002qubit,goto2004multiqubit,zheng2005nongeometric,lacour2006arbitrary}. However, such adiabatic methods are based on the adiabatic theorem of quantum mechanics~\cite{messiah2014quantum} and can thereby be time-consuming due to the necessity to fulfill the adiabatic condition
\begin{equation}
	\left | \frac{\langle \phi_m(t)|\dot{H}(t)|\phi_n(t) \rangle}{[E_n(t) - E_m(t)]^2} \right | \ll 1,
	\label{eq:adiabatic_condition}
\end{equation}
where $|\phi_m(t)\rangle$ and $|\phi_n(t)\rangle$ denote the eigenstates of the time-dependent Hamiltonian $H(t)$ with the corresponding eigenenergies $E_m(t)$ and $E_n(t)$, whereas $\dot{H}(t)$ represents the time derivative of $H(t)$. This condition requires a slow driving in order to ensure that the system remains in an instantaneous eigenstate throughout the process. Therefore, considerable attention has been focused on methods to speed up adiabatic processes, both theoretically~\cite{chen2010shortcut,giannelli2014superadiabatic,liang2016superadiabatic,baksic2016speeding,petiziol2020superadiabatic,stefanatos2020speeding,delcampo2019focus,gueryodelin2019review} and experimentally in various platforms~\cite{schaff2011coldgases,bason2012boseeinstein,malossi2013boseeinstein,zhang2013nvexperiment,xu2019breaking,li2018superconducting,vepsalainen2019superconducting,zheng2022superconducting,zheng2022arxiv}.

Remarkably, the possibility of a quantum adiabatic evolution even in the presence of vanishing energy gaps has been demonstrated theoretically~\cite{wang2016necessary} and experimentally~\cite{xu2019breaking}, respectively. The proposed method is a protocol employing discrete jumps along the evolution path of the control parameters to realize quantum adiabatic processes at unlimited rates in a two-level system. Such a jump protocol enables a rapid evolution that can even avoid path points where the eigenstates of the Hamiltonian are not experimentally feasible~\cite{wang2016necessary}.

Among the platforms for practical implementations of quantum technologies at ambient conditions, the negatively charged nitrogen-vacancy (NV) center in diamond~\cite{gruber1997scanning,rondin2014magnetometry} represents an appealing and promising candidate, due to its long coherence time at room temperature and well developed coherent control techniques~\cite{jelezko2004observation,bar2013solid,balasubramanian2009ultralong,chu2015all,shu2018observation,tian2020quantum,cao2020pra}. Therefore, the NV system has a great number of applications in quantum information processing~\cite{wrachtrup2006processing,dolde2013room,cai2013large,xiang2013hybrid,kurizki2015quantum,yu2020experimental,yu2022quantum,yu2021experimental} and quantum computing~\cite{neumann2008multipartite,zu2014experimental}. Besides other favorable properties, the energy-level structure of the NV center~\cite{jelezko2006single,doherty2013nitrogen} makes it a quantum-sensing platform for temperature~\cite{kucsko2013nanometre,Acosta2010prl,neumann2013nl,plakhotnik2014nl}, strain~\cite{knauer2020situ,trusheim2016njp,kehayias2019prb}, electric~\cite{dolde2011electric} and magnetic fields~\cite{maze2008nanoscale,balasubramanian2008nanoscale,welter2022magnetometry}, as well as a hybrid sensor~\cite{cai2014nc,wang2018prx,liu2020nsr}.

In this work, we utilize a single NV center to experimentally implement a jump protocol in a three-level system~\cite{liu2022arxiv} to speed up the adiabatic state transfer by jumping along the geodesics in the three-level system. To demonstrate the speed up, we compare the population transfer to the well-established STIRAP. We find that the jump scheme exhibits a high transfer efficiency at appreciably shorter times. Furthermore, we demonstrate that it is robust against environmental magnetic noise, which illustrates the feasibility of the jump protocol in realistic environments.

\section{Jump protocol}
The conventional quantum adiabatic theorem requires the evolution of the quantum systems to be subject to slowly varying Hamiltonians, which imposes a speed limit on quantum adiabatic methods. On the other hand, due to the limited coherence time of quantum systems, a fast coherent control is desired in order to attain high operation fidelities. Therefore, intensive work has been done to develop new methods to improve the efficiency of quantum adiabatic processes. In Ref.~\cite{wang2016necessary}, a method was proposed in which the adiabatic evolution is decomposed into a product of gauge-invariant unitary operators and a necessary and sufficient condition for adiabaticity is provided. This, in turn, has prompted a new scheme utilizing parametrized pulse sequences to improve the speed of the adiabatic evolution~\cite{xu2019breaking}. 

In this Letter, we consider a system comprised of the three states $|-1\rangle$, $|0\rangle$, and $|+1\rangle$ where there is no direct transition between the two states $|-1\rangle$ and $|+1\rangle$ [see Fig.~\ref{fig:pulse_seq}(a)] and which is described by the parametrized Hamiltonian $H(\theta)$. Here, let the parameter $\theta=\theta(t)$ be a dimensionless time-dependent monotonic function with the initial condition $\theta(t=0)=0$. Our aim is to realize an adiabatic population transfer between $|-1\rangle$ and $|+1\rangle$ by variation of the parameter $\theta$. We denote the instantaneous eigenstates and eigenenergies of $H(\theta)$  by  $\{|n(\theta)\rangle\}$ and $\{E_n(\theta)\}$, respectively.  

According to Ref.~\cite{liu2022arxiv}, for a total evolution time $T$ the time-evolution operator $U(T)$ generated by $H(\theta)$, from $t=0$ to $t=T$, can be written as the product
\begin{equation}
	U(T) = U_{{\rm adia}} U_{\text{Dia}} (T), \label{eq:UUU}
\end{equation}
where $U_{{\rm adia}}$ is the time-evolution operator of the ideal adiabatic process, whereas the undesirable nonadiabatic correction reads
\begin{equation}
	U_{\text{Dia}} (T) = \mathcal{P} \exp\left[i\int_{0}^{\theta_T} X(\theta) d\theta\right],
	\label{eq:nonadiabatic_correction}
\end{equation}
with the shorthand $\theta_T\equiv \theta(t=T)$. Furthermore, $\mathcal{P}$ represents the path-ordering operation~\cite{liu2022arxiv,Carollo2006} with respect to the parameter $\theta$, in the same fashion as the well-known time-ordering operator~\cite{rivas}. Lastly, we have also defined the quantity 
\begin{equation}
X(\theta)=\sum_{n,m} \xi_{n,m}(\theta) G_{n,m}(\theta)\label{eq:Xxig}
\end{equation}
in terms of
\begin{gather}
	\xi_{n,m}(\theta) = \exp\{i[\varphi_n(\theta) - \varphi_m(\theta)]\},\\
	G_{n,m}(\theta) = \langle n(\theta)|i\frac{d}{d \theta}|m(\theta)\rangle |n(0)\rangle\langle m(0)| \label{eq:Gnm}
\end{gather}
with the dynamical phases $\varphi_j(\theta)=\int_0^tE_j(t')dt'$. Thus, from Eq.~(\ref{eq:UUU}) we see that if $U_{\mathrm{Dia}}(T)$ equals the identity operator $I$, all nonadiabatic effects are eliminated.

We would like to realize an adiabatic transfer by a dark state, $| d(\theta)\rangle$, which is absent in the Hamiltonian $H(\theta)$ and corresponds to an eigenstate with a zero eigenvalue.  If we choose $|-1\rangle$ as the initial dark state, the initial Hamiltonian takes the form	$H(\theta=0)=\Omega|0\rangle\langle -1|+\mathrm{H.c.}+\Delta|0\rangle\langle 0|$, 
where $\Omega$ is the amplitude and $\Delta$ is the detuning.  The value of $\Delta$ can be used to optimize the performance when there are dissipation and/or decoherence on both the intermediate state $|0\rangle$ and the qubit sub-space $\{|\pm1\rangle\}$~\cite{liu2022arxiv}. For simplicity, here we consider $\Delta=0$, as it allows for a faster control, such that possible detrimental effects of decoherence on $\{|\pm1\rangle\}$ can be minimized. 
The eigenstates of $H(\theta=0)$ are $|d(0)\rangle = |-1\rangle$ and $|\mu_\pm(0)\rangle=(|+1\rangle \pm  |0\rangle)/2$. 

As in STIRAP, choosing the dark state to have the form
\begin{equation}
| d(\theta)\rangle=\sin(\theta)|+1\rangle+\cos(\theta)|-1\rangle,
\end{equation}
we find that $|d(\theta)\rangle = \exp(-i G \theta)| d(0)\rangle$, with the constant generator $G = i|-1\rangle \langle +1| -i |+1\rangle \langle -1|$. We note that 
$dG/d\theta=0$, which results in the important property that the $G_{n,m}$ in Eq.~(\ref{eq:Gnm}) are all constant~\cite{liu2022arxiv} if we use the eigenstates $\{|n (\theta)\rangle\}= \{|d(\theta)\rangle, |\mu_\pm(\theta)\rangle\}$ of the Hamiltonian
\begin{align}
	H(\theta) &= \exp(-i G \theta) H(0) \exp(i G \theta), \nonumber \\
	&= \Omega|\mu_+(\theta)\rangle\langle \mu_+(\theta)| - \Omega|\mu_-(\theta)\rangle\langle \mu_-(\theta)|,\nonumber \\ 
	&= \Omega [\cos(\theta)|+1\rangle-\sin(\theta)|-1\rangle]\langle 0|+ \mathrm{H.c.},
	\label{eq:hamiltonian_theory}
\end{align}
where $|\mu_{\pm}(\theta)\rangle = \exp(-i G \theta)|\mu_{\pm}(0)\rangle $. We can derive the relation~\cite{liu2022arxiv} 
 \begin{equation}
G_{n,m} = \langle n(0) | G | m(0)\rangle | n(0)\rangle  \langle m(0)|, \label{eq:GnmG}
\end{equation}
which also shows that Eq.~(\ref{eq:Gnm}) becomes a constant.

Now, we apply $H(\theta)$ for a duration $\tau=\pi/\Omega$ at $N$ equally-spaced points $\theta = \theta_1,\theta_2,\ldots, \theta_N$ between $\theta_0= 0$ and $\theta_T=\theta_{N+1}\equiv\pi/2$. Specifically, we choose the $N$ parameter values $\theta_j$ ($j=1,\dots,N$) to be
\begin{equation}
	\theta_j = \frac{2j - 1}{N}\frac{\pi}{4}.  \label{eq:theta_j}
\end{equation}
The time-evolution operator thereby becomes
$U(T)=\exp[-iH(\theta_N)\tau]\exp[-i H(\theta_{N-1})\tau]\dots\exp[-i H(\theta_1)\tau]$. Additionally,
we find $\xi_{d,\mu_+}(\theta)=\xi_{d,\mu_-}(\theta)=F(\theta)$, with 
\begin{equation}
F(\theta)=(-1)^j, \label{eq:Ft}
\end{equation}
$\theta\in[\theta_j,\theta_{j+1})$, since applying $H(\theta)$ in Eq.~(\ref{eq:hamiltonian_theory}) with a fixed value of $\theta$ for a time $\tau=\pi/\Omega$ yields a shift of $\pm\pi$ between the dynamic phases of $|\mu_\pm(\theta)\rangle$ and $| d(\theta)\rangle$.  On the other hand, evaluating Eq.~(\ref{eq:GnmG}) shows $G_{n,n}=G_{\mu_+,\mu_-}=G_{\mu_-,\mu_+}=0$. 
Equation~(\ref{eq:nonadiabatic_correction}) thus takes the form
\begin{equation}
U_{\text{Dia}}(T)=\exp\left[i\left(\sum_{n=\mu_{\pm}}G_{n,d}+\mathrm{H.c.}\right)\int_0^{\theta_T} F(\theta) d\theta\right], 
\end{equation}
which finally reveals $U_{\text{Dia}}(T)=I$
by using Eqs.~\eqref{eq:theta_j} and~\eqref{eq:Ft}. This means our choice of parameter jumps realizes the desired adiabatic evolution $U(T)=U_\mathrm{adia}$ in the finite time $T=N\tau$.

\section{Performance of the jump protocol}
To demonstrate the high population transfer efficiency that can be achieved with this jump protocol, we experimentally compare it with a traditional STIRAP protocol in an NV-center system. The STIRAP protocol utilizes two Raman pulses to realize the population transfer between two states under the adiabatic condition. In our case, we use Raman pulses with a Gaussian envelope. Both optical and microwave fields can be used for coherent control in the NV center system, we implement the STIRAP protocol by microwave fields whose amplitude and phase are allowed to be better controlled. 

In the experiment, we choose the ground-state triplet of a single NV center as the three-level system, respectively identifying $|0\rangle$ and $|\pm1\rangle$ with the $m_s=0$ and $m_s=\pm1$ states. The degeneracy of the $m_s=\pm1$ states is lifted by an external magnetic field aligned with the NV-center axis. An acousto-optical modulator (AOM) is used to control the 532-nm green laser which is employed for the optical ground-state initialization and the readout of the spin state. In order to manipulate the spin states, we apply two microwave driving fields which are resonant with the $|0\rangle\leftrightarrow|\pm1\rangle$ transitions through an arbitrary waveform generator (AWG). The Rabi frequencies of the two driving fields are denoted by $\Omega_\pm$.

In the interaction picture with respect to the NV-center ground-state Hamiltonian, the three-level system is described by the Hamiltonian in Eq.~\eqref{eq:hamiltonian_theory}. The parameter $\Omega$ and the Rabi frequencies $\Omega_\pm$ of the microwave driving fields are related by $\Omega_-=2\Omega\sin(\theta)$ and $\Omega_+=2\Omega\cos(\theta)$. By changing $\Omega_-$ and $\Omega_+$ via the AWG, while keeping $\Omega_-^2+\Omega_+^2$ constant, we are able to tune the value of $\theta$. In the experiment, we set $\Omega/2\pi=4$~MHz and since the duration $\tau$ of each control step with $\theta_j$ in the jump protocol should be equal to $\pi/\Omega$ in order to fulfill the adiabatic condition we use $\tau=0.125\ \mu$s.

\begin{figure}[tb]
	\centering
	\includegraphics[width=8.6cm]{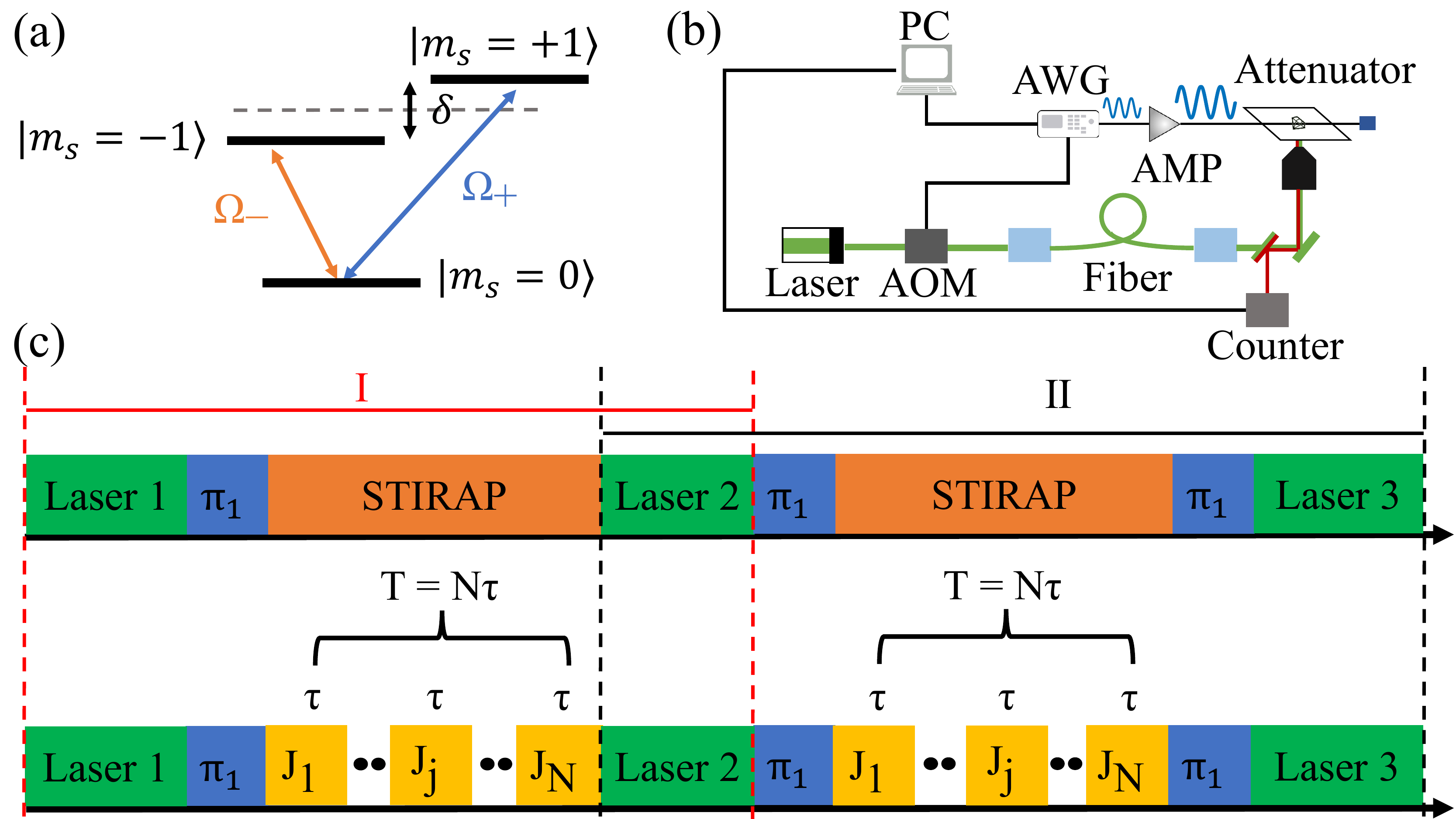}
	\caption{(a) The energy difference $\delta=\gamma B$ between $m_s=\pm 1$ arises from the Zeeman effect. (b) The experiment setup for the coherent control of the NV center.
	(c) Upper part: pulse sequence for the STIRAP protocol used as comparison. Lower part: pulse sequence for the jump protocol. Parts I and II measure the population of $|0\rangle$ and $|-1\rangle$, respectively. Laser initialization and readout are the same for both schemes. $\pi_1$ corresponds to $\pi$ pulses on the transition $|0\rangle\leftrightarrow|-1\rangle$. In the jump protocol, there are $N$ successive control pulses of length $\tau$, represented by $J_j$, that correspond to the piecewise constant values $\theta_j$.}
	\label{fig:pulse_seq}
\end{figure}

Figure~\ref{fig:pulse_seq} shows a schematic of the experimental setup. Figure.~\ref{fig:pulse_seq}(a) depicts the energy-level structure of the NV ground state. Here, the energy splitting $\delta$, with the gyromagnetic ratio $\gamma/2\pi=2.8$~MHz/G, between the states $|m_s=\pm1\rangle$ stems from the Zeeman effect due to the applied magnetic field $B$. The explicit setup is sketched in Fig.~\ref{fig:pulse_seq}(b) and Fig.~\ref{fig:pulse_seq}(c) shows the pulse sequences we apply. The upper part depicts the sequence for the STIRAP protocol we will use for comparison, whereas the lower one represents the sequence for the jump protocol. 
Since we aim to measure the transfer efficiency in the three-level system, the pulses sequences consists of two parts, namely parts I and II, from which we can infer the transfer efficiency given by the population of $|+1\rangle$. In detail, both sequences begin with laser 1 to initialize the state in $|0\rangle$ followed by a $\pi$ pulse on the $|0\rangle\leftrightarrow|-1\rangle$ transition. Then the actual protocols of length $T$ are applied, where the building blocks $J_j$ represent the $N$ control steps of $\theta$ in the jump protocol. Laser 2 reads out the population of $|0\rangle$ at the end of part I and, at the same time, re-initializes the electron spin for part II, which differs only by an additional $\pi$ pulse before read-out, such that the population of $|-1\rangle$ is measured.

\begin{figure}[tb]
	\centering
	\flushleft\normalsize{(a)}\hspace{-2.4ex}\includegraphics[width=1\linewidth]{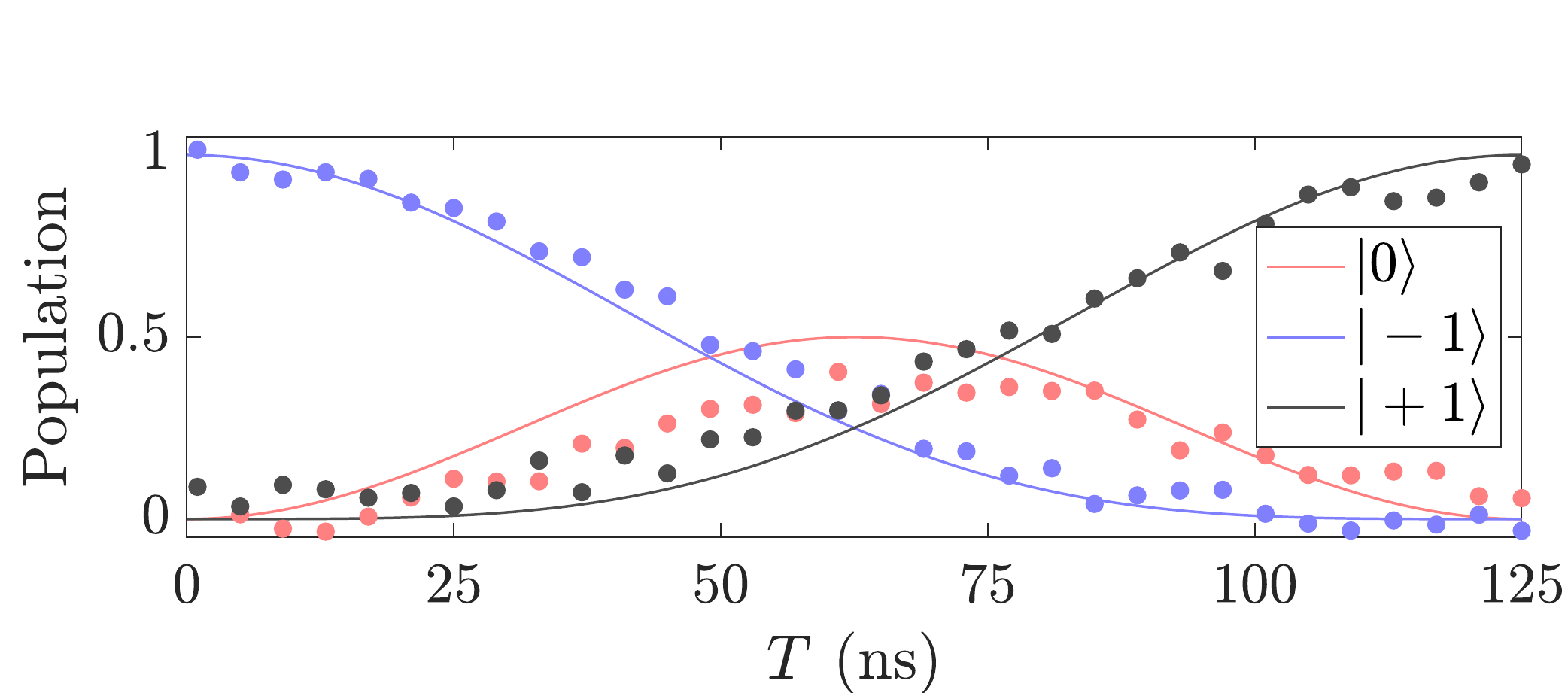}\vspace{-3ex}
    \flushleft\normalsize{(b)}\hspace{-2.4ex}\includegraphics[width=\linewidth]{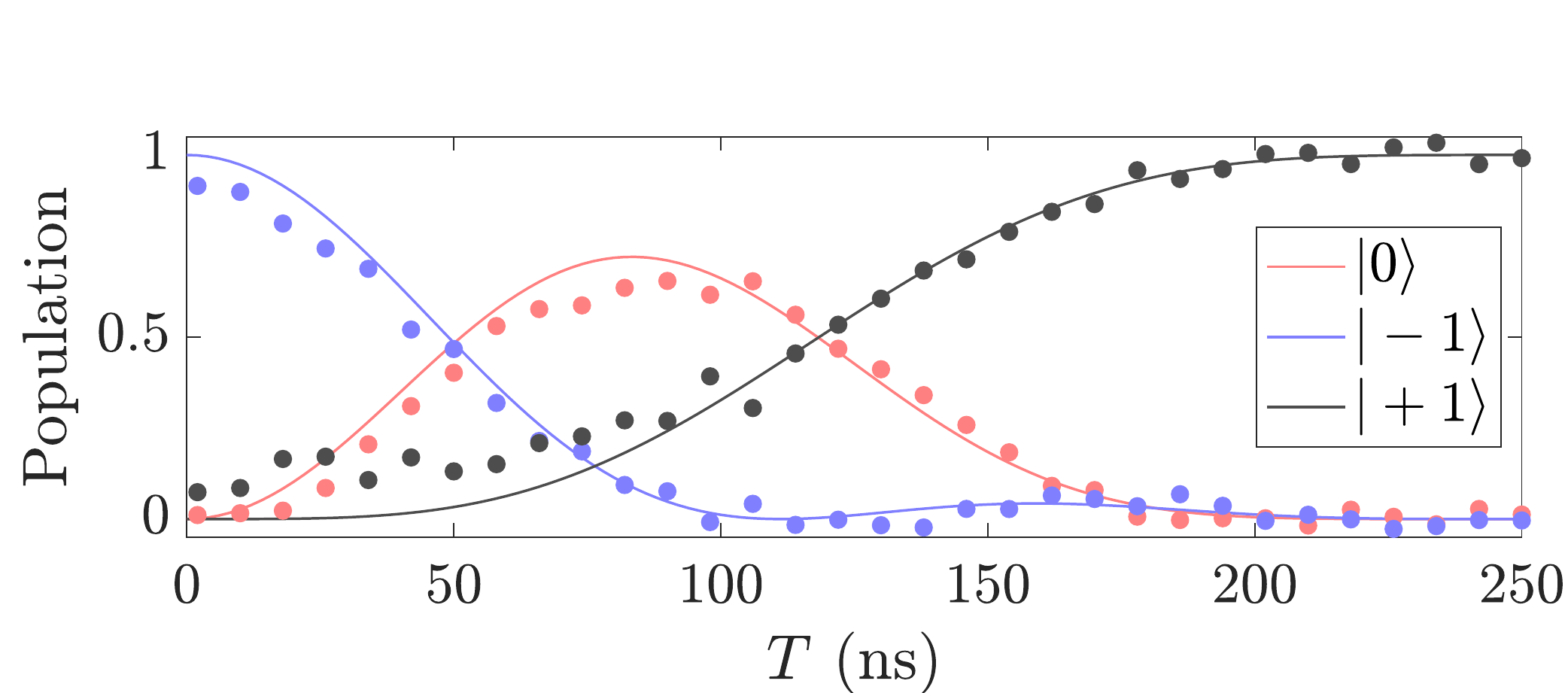}\vspace{-3ex}
    \flushleft\normalsize{(c)}\hspace{-2.4ex}\includegraphics[width=\linewidth]{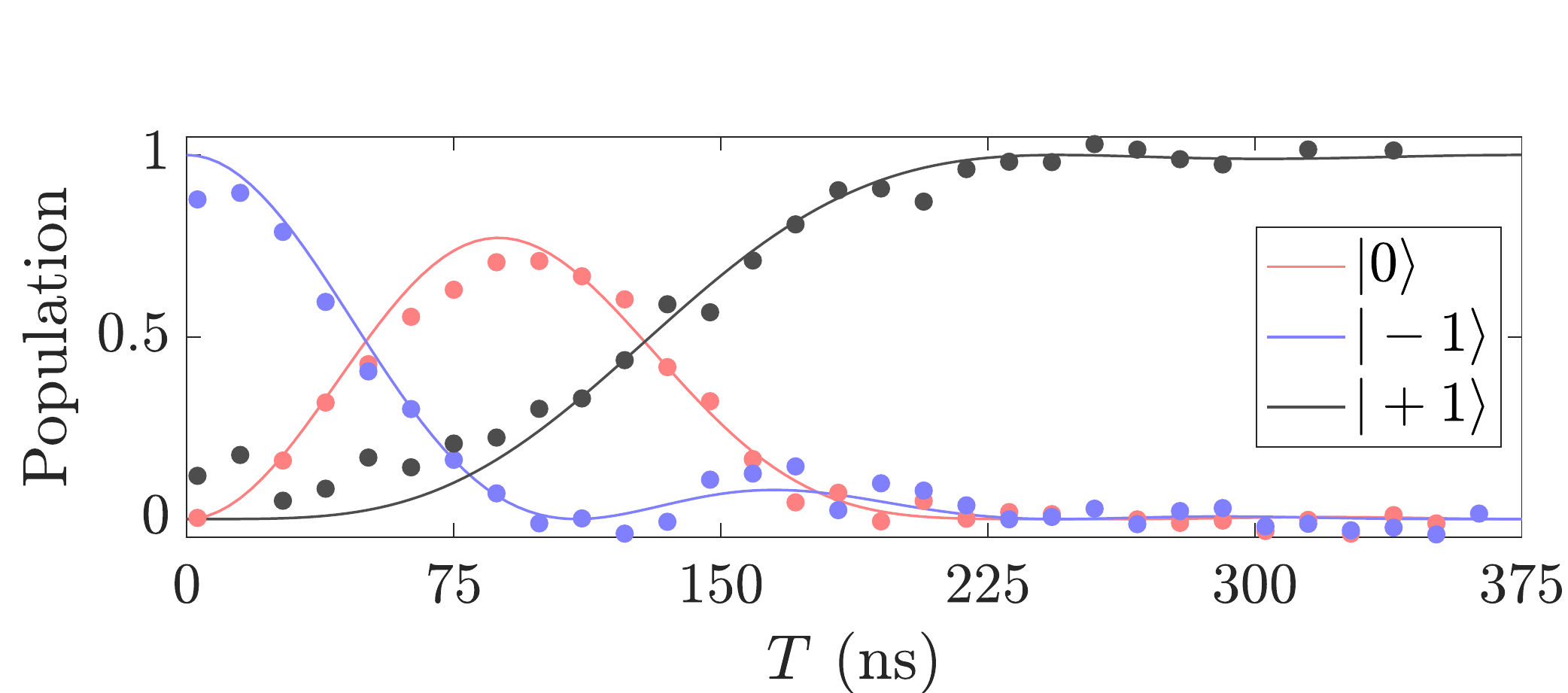}\vspace{-3ex}
    \flushleft\normalsize{(d)}\hspace{-2.4ex}\includegraphics[width=\linewidth]{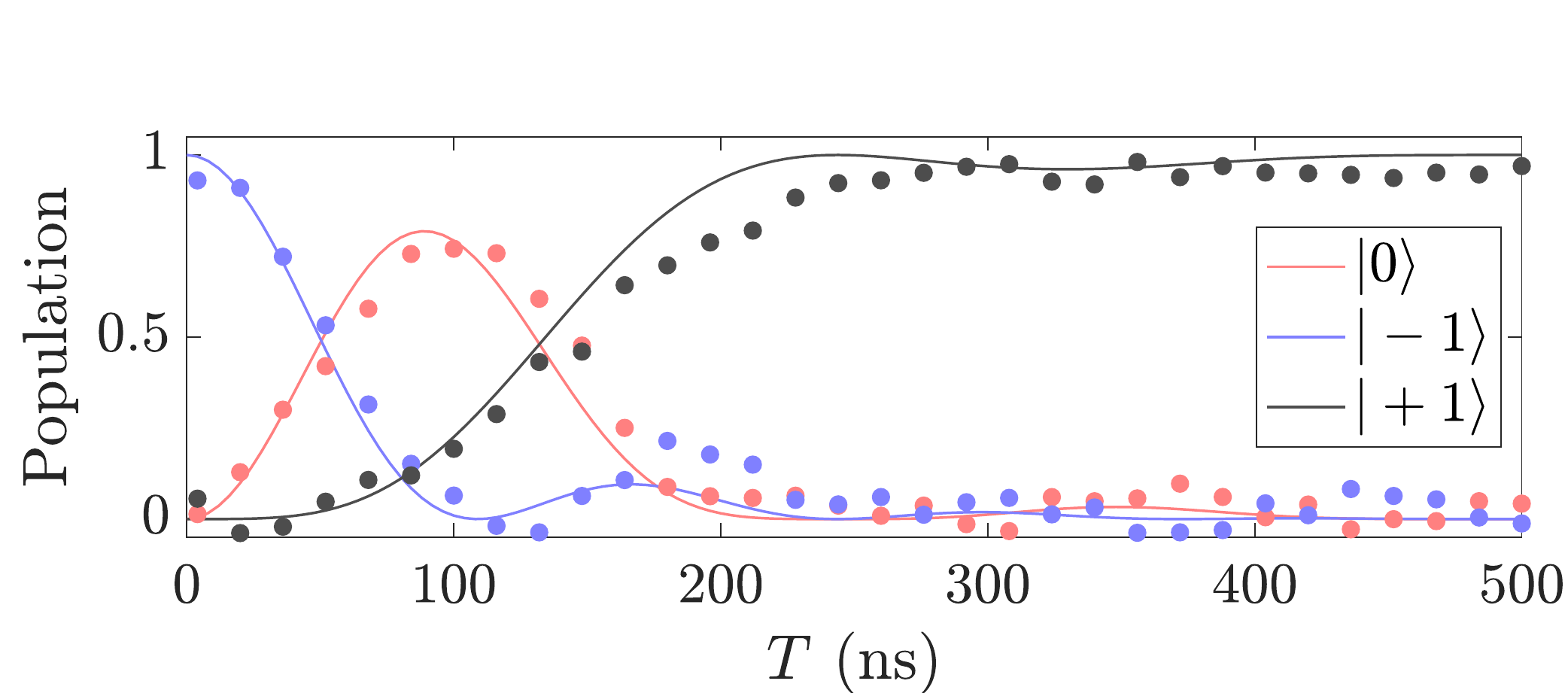}
	\caption{Dependence of the populations of the three-level system on the total evolution time $T$ in the jump protocol. (a)-(d) respectively correspond to $N=1,2,3,4$ control pulses with $\Omega/2\pi=4$ MHz. For every value of $T$ each of the $N$ pulses has the length $T/N$. Markers show experimental results and solid lines are numerical simulations.}
	\label{fig:Jump_N_1_2_4}
\end{figure}
To demonstrate the applicability of the jump protocol we measure the population transfer from $|-1\rangle$ to $|+1\rangle$ by tracking the population of all three state of the system in dependence of the total evolution time $T$ for different numbers of pulses. For a given value of $T$ each of the $N$ pulses thereby has the length $T/N$. This is shown in Fig.~\ref{fig:Jump_N_1_2_4}, where Figs.~\ref{fig:Jump_N_1_2_4}(a)--\ref{fig:Jump_N_1_2_4}(d) respectively depict the four pulse numbers $N=1,2,3,4$, corresponding to the final values of the evolution times $T=125,250,375,500$ ns. Here, the markers are experiment results, whereas solid lines show numerical simulation. We find a good agreement and the results show that the duration of 125 ns for a single-pulse protocol leads to a full population transfer. It can also be seen that the population in the target state $|+1\rangle$ with a value close to unity has a wider range at the end of the protocol for larger $N$, which indicates the robustness of multiple-pulse protocols with respect to amplitude imperfections.

We now compare the performance of the jump protocol to the well-established STIRAP protocol. Details on the STIRAP measurements are given in the Appendix. For better comparison, in Fig.~\ref{fig:Jump_Transfer_Efficiency}(a), we show the transfer efficiency, as given by the population of $|+1\rangle$, for $N=1,2,3,4$ [corresponding to the black data from Figs.~\ref{fig:Jump_N_1_2_4}(a)--\ref{fig:Jump_N_1_2_4}(d)].
\begin{figure}[tb]
	\centering
	\flushleft\normalsize{(a)}\hspace{-2.4ex}\includegraphics[width=1\linewidth]{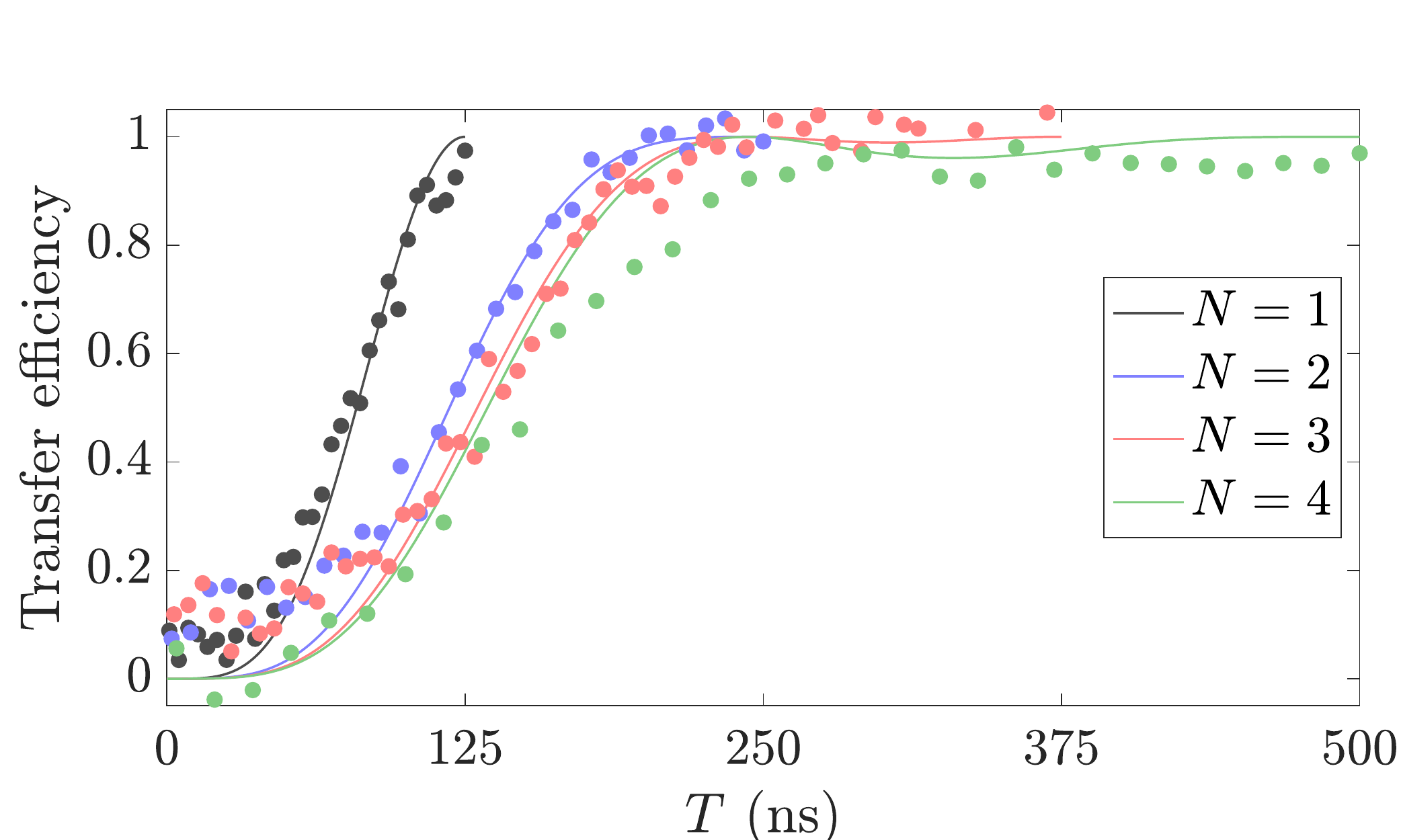}\vspace{-3ex}
    \flushleft\normalsize{(b)}\hspace{-2.4ex}\includegraphics[width=\linewidth]{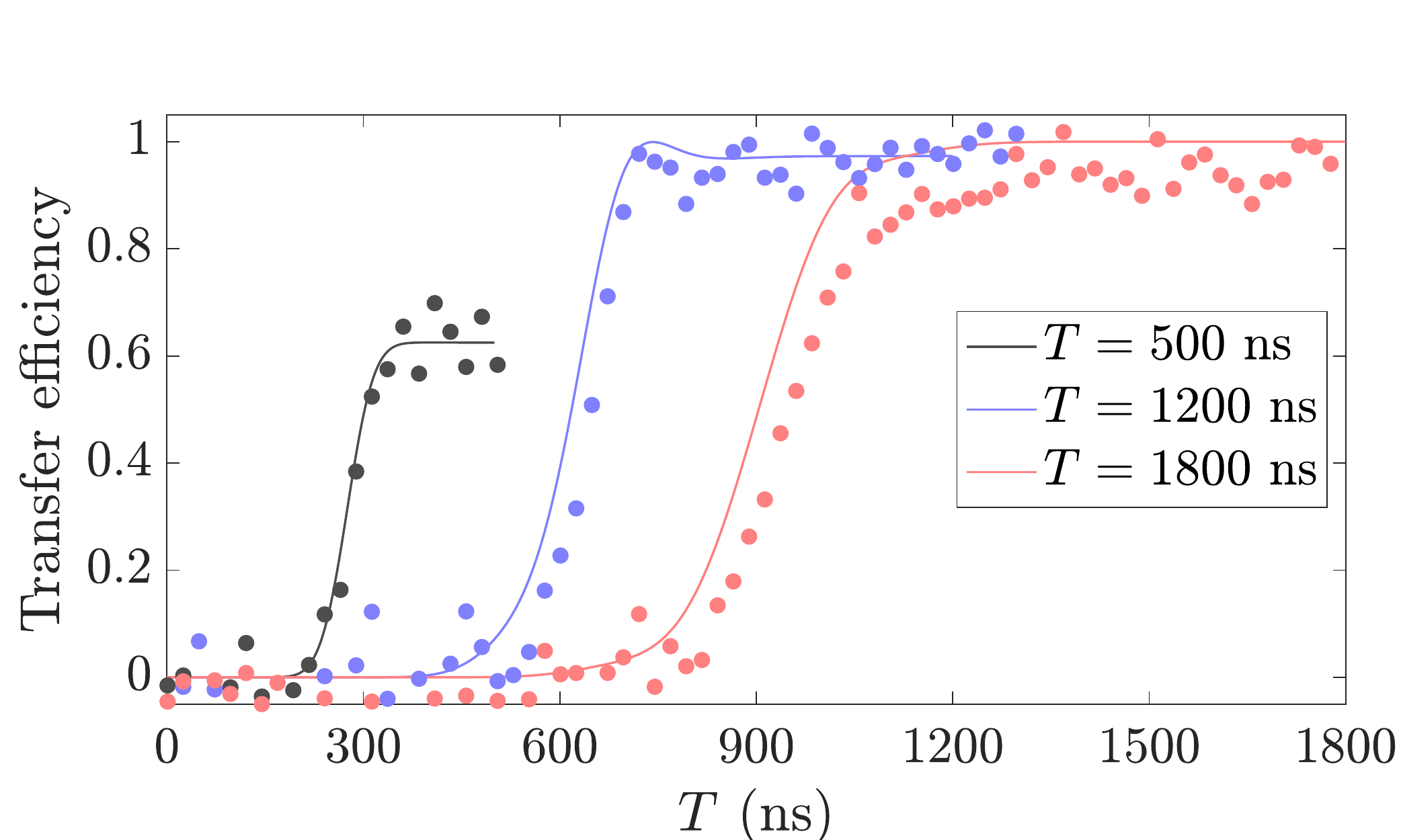}
	\caption{Comparison of the transfer efficiency between the jump protocol and STIRAP. (a) Different number of control pulses, $N=1,2,3,4$, of the jump protocol. (b) STIRAP for different final evolution times, $T=500,1200,1800$ ns. Markers and solid lines represent experimental data and numerical simulations, respectively.}
	\label{fig:Jump_Transfer_Efficiency}
\end{figure}
Figure~\ref{fig:Jump_Transfer_Efficiency}(b), on the other hand, shows the results of the STIRAP protocol for three different evolution times, namely $T=500,1200,1800$ ns. For the shortest evolution time, which is the longest we use for the jump protocol, the transfer efficiency only reaches around 60\%. With the parameters we chose, i.e., a maximum amplitude of the Raman control pulses of $2\Omega$, the time required for a complete population transfer is well above 1000 ns and thereby appreciably longer than the one required in the jump protocol. In both Figs.~\ref{fig:Jump_Transfer_Efficiency}(a) and~\ref{fig:Jump_Transfer_Efficiency}(b), markers and solid lines represent experimental and numerical simulation results, respectively.

\section{Robustness against noise}
Finally, we demonstrate the robustness of the jump protocol against environmental noise. In the solid-state spin system, the main source of noise is the surrounding bath of nuclear spins which can be described by an effective magnetic field~\cite{dobrovitski2009spinnoise}. To investigate the detrimental effects of static magnetic field noise, we artificially add a detuning $\pm\Delta$ between the microwave driving fields and the transitions $|0\rangle\leftrightarrow|\mp1\rangle$, while keeping the remaining parameters unchanged. We then compare the robustness of the transfer efficiency of the jump protocol to the one of the STIRAP. In this case, the Hamiltonian for the jump and STIRAP protocol respectively read
\begin{equation}
	H_I=
	\begin{pmatrix}
	0 & \Omega \cos \phi & \Omega \sin \phi \\
	\Omega \cos \phi & \Delta & 0\\
	\Omega \sin \phi & 0 & -\Delta\\
	\end{pmatrix},
 \end{equation}
 \begin{equation}
	H_{\text{STIRAP}} = \frac{1}{2} 
	\begin{pmatrix}
	0 & \Omega_P (t) & \Omega_S (t)\\
	\Omega_P (t) & \Delta & 0\\
	\Omega_S (t) & 0 & -\Delta\\
	\end{pmatrix}.
\end{equation}

\begin{figure}[tb]
	\centering
	\includegraphics[width=\linewidth]{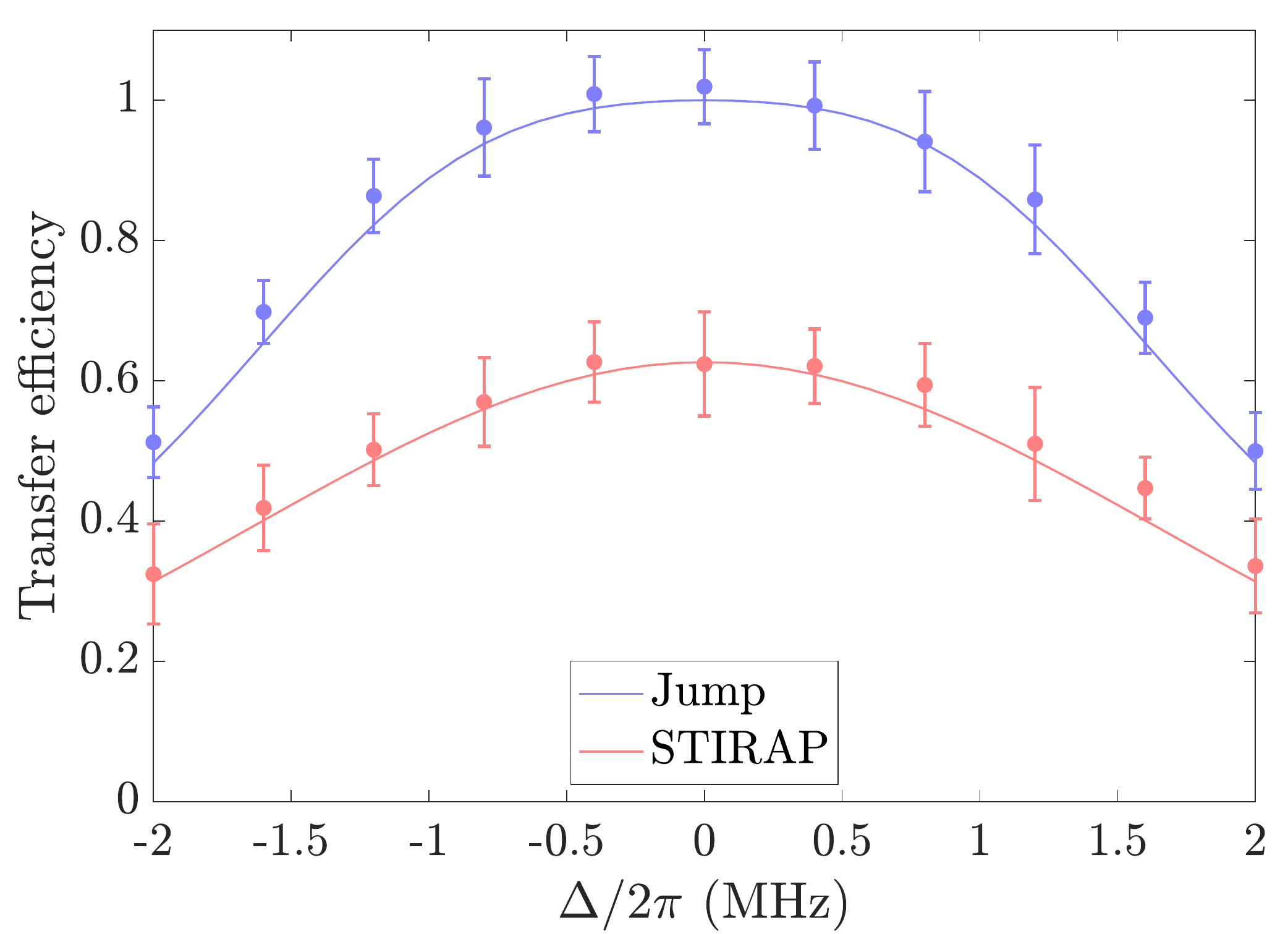}
	\caption{Comparison of the robustness against frequency detuning between the jump protocol (blue) and STIRAP (red). The evolution time is $T = 500$ ns for both cases ($N=4$ in the jump protocol). Markers with error bars and solid lines represent experimental data and numerical simulations, respectively.}
	\label{fig:Robustness}
\end{figure}
\noindent For the comparison of the two schemes, we again set the frequency $\Omega/2\pi=4$ MHz and chose $N=4$ as the number of control pulses. This implies a total evolution time of $T=500$~ns, which we also set as the  total duration of the STIRAP protocol. The maximal amplitudes of the pump and Stokes pulses, $\Omega_P(t)$ and $\Omega_S(t)$, in the STIRAP are both given by $2\Omega$. The experiment results for varying the detuning $\Delta/2\pi$ from $-2$~MHz through $2$ MHz are shown in Fig.~\ref{fig:Robustness}. Solid lines represent simulation results whereas the markers with error bars show experimental results. As one sees, for the same evolution time, the jump scheme exhibits not only a higher transfer efficiency throughout the whole range of the detuning, but also a much flatter central part, indicating a better robustness than the STIRAP.

\section{Conclusions}
We have demonstrated jump scheme for population transfer in a three-level system realized in single NV-center spin in diamond. By comparing the protocol to a traditional STIRAP scheme, we have shown its significant advantages. In particular, the state-transfer speed and efficiency is greatly improved, i.e., much shorter times are required to achieve a high-fidelity population transfer. Furthermore, we have also compared the impact of noise on the two schemes under the same evolution time and established an improved robustness of the jump protocol. The developed method is thereby a promising candidate for practical applications in quantum control, quantum sensing, quantum information processing, and even chemical interaction control.

\begin{acknowledgments}
We thank Dongxiao Li for discussions and help during the initial stage of manuscript preparation. R. B. is grateful to S. Schein for helpful comments. This work is supported by the National Natural Science Foundation of China (Grants No.~12161141011, No.~11874024, and No.~12074131), the National Key R$\&$D Program of China (Grant No. 2018YFA0306600), the Shanghai Key Laboratory of Magnetic Resonance (East China Normal University), and the Natural Science Foundation of Guangdong Province (Grant No. 2021A1515012030). 
\end{acknowledgments}

\appendix
 \section{STIRAP}
To achieve the population transfer through an auxiliary state in the STIRAP, one utilizes two Raman control pulses with the envelopes $\Omega_S(t)$ and $\Omega_P(t)$, called the Stokes and pump pulses. The NV-center ground state consists of three sublevels, $m_s = 0, \pm 1$. The states $|0\rangle$ and $|\pm 1\rangle$ are coupled, whereas $|-1\rangle \leftrightarrow |+1\rangle$ is a forbidden transition. Thus, the primary aim of STIRAP is to drive the $|-1\rangle \leftrightarrow |+1\rangle$ transition  using $|0\rangle$ as an auxiliary state without or with only minimal population loss to the latter state. The two Raman pulses are implemented with microwave fields in our experiment.

The NV-center spin is prepared in the state $|-1\rangle$ by applying a $\pi$-pulse on the $|0\rangle\leftrightarrow|-1\rangle$ transition after optical initialization. We then apply two resonant microwave control pulses with the envelopes $\Omega_S(t)$ and $\Omega_P(t)$ for the population transfer to $|+1\rangle$. The Stokes pulse $\Omega_S(t)$ is applied first to drive the transition $|0\rangle\leftrightarrow|+1\rangle$ and is followed by the pump pulse $\Omega_P (t)$ which drives the transition $|0\rangle\leftrightarrow|-1\rangle$. In the interaction picture with respect to the ground-state Hamiltonian, the Hamiltonian of STIRAP scheme is given by
\begin{equation}
	H_\mathrm{STIRAP}=\frac{1}{2} \begin{pmatrix}
		0 & \Omega_P (t) & \Omega_S (t)\\
		\Omega_P (t) & 0 & 0\\
		\Omega_S (t) & 0 & 0\\
	\end{pmatrix}
\end{equation}
in the basis $\{|0\rangle, |-1\rangle, |+1\rangle\}$.

\begin{figure}[tb]
	\centering
	\includegraphics[width=\linewidth]{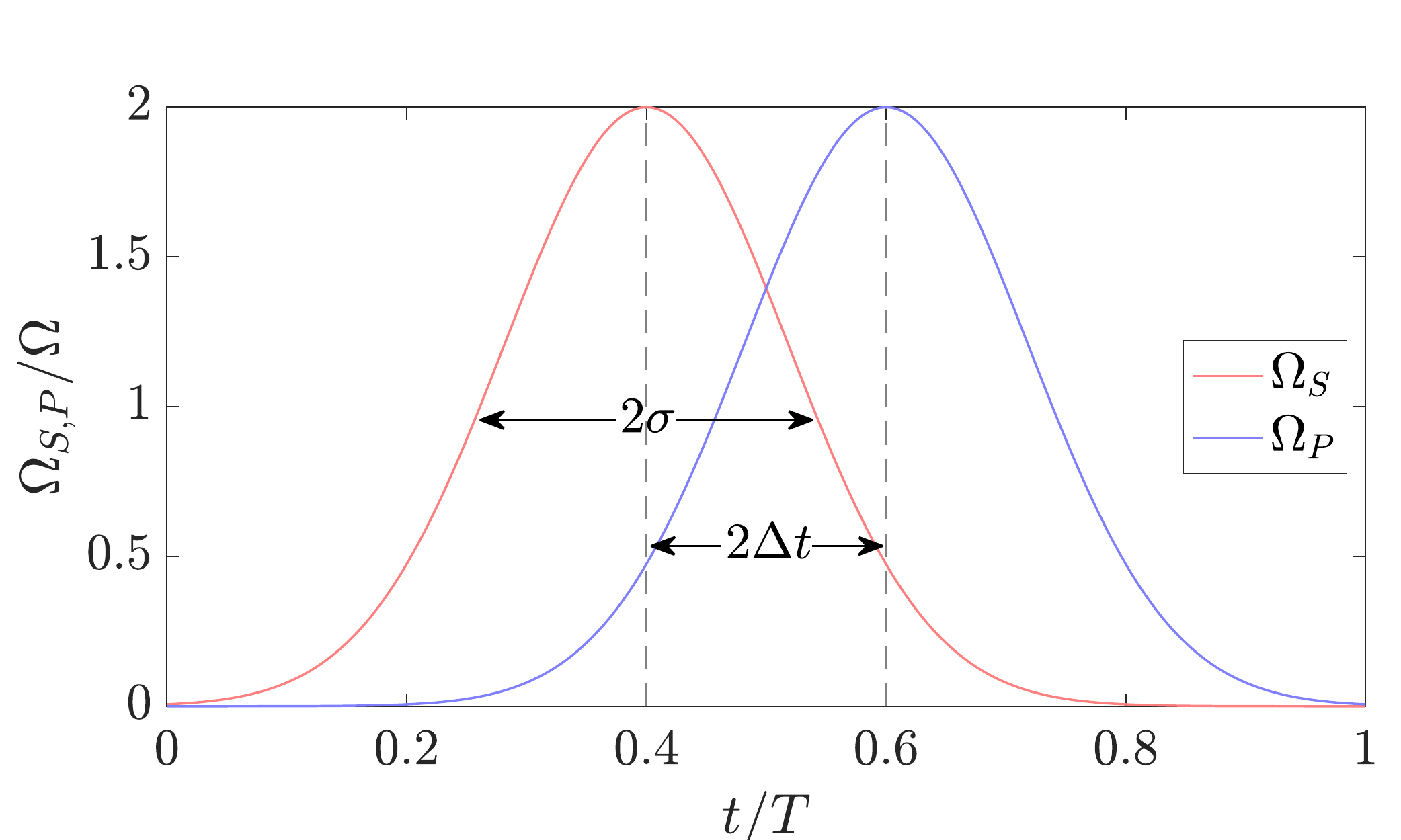}
	\caption{The two Raman pulses in the STIRAP scheme. The red and blue lines correspond to the control pulses $\Omega_S(t)$ and $\Omega_P(t)$, respectively. Both pulses have a Gaussian envelope with the full-width at half-maximum $2\sigma$ and the peak value $2\Omega$. They partially overlap and the delay between them is $2\Delta t$.}
	\label{fig:stirap_pulse_shape}
\end{figure}

The two Raman control pulses are applied adiabatically and partially overlap to completely transfer the population, see
Fig.~\ref{fig:stirap_pulse_shape}. 
They have Gaussian shapes and for a total evolution time $T$ they are centered at $T/2\pm\Delta t$, i.e., the second pulse is delayed by $2\Delta t$. Furthermore, their full-width at half-maximum is $2\sigma$ and their maximal value is $2\Omega$. This means they have the form
$\Omega_S (t)=2\Omega\exp[-(t-T/2+\Delta t)^2/\sigma^2]$ and 
$ \Omega_P (t)=2\Omega\exp[-(t-T/2-\Delta t)^2/\sigma^2]$. In our experiment, we set $\Omega/2\pi=4$ MHz, $\Delta t=T/10$, and $\sigma=T/6$. 
\begin{figure}[tb]
	\centering    \flushleft\normalsize{(a)}\hspace{-2.4ex}\includegraphics[width=1\linewidth]{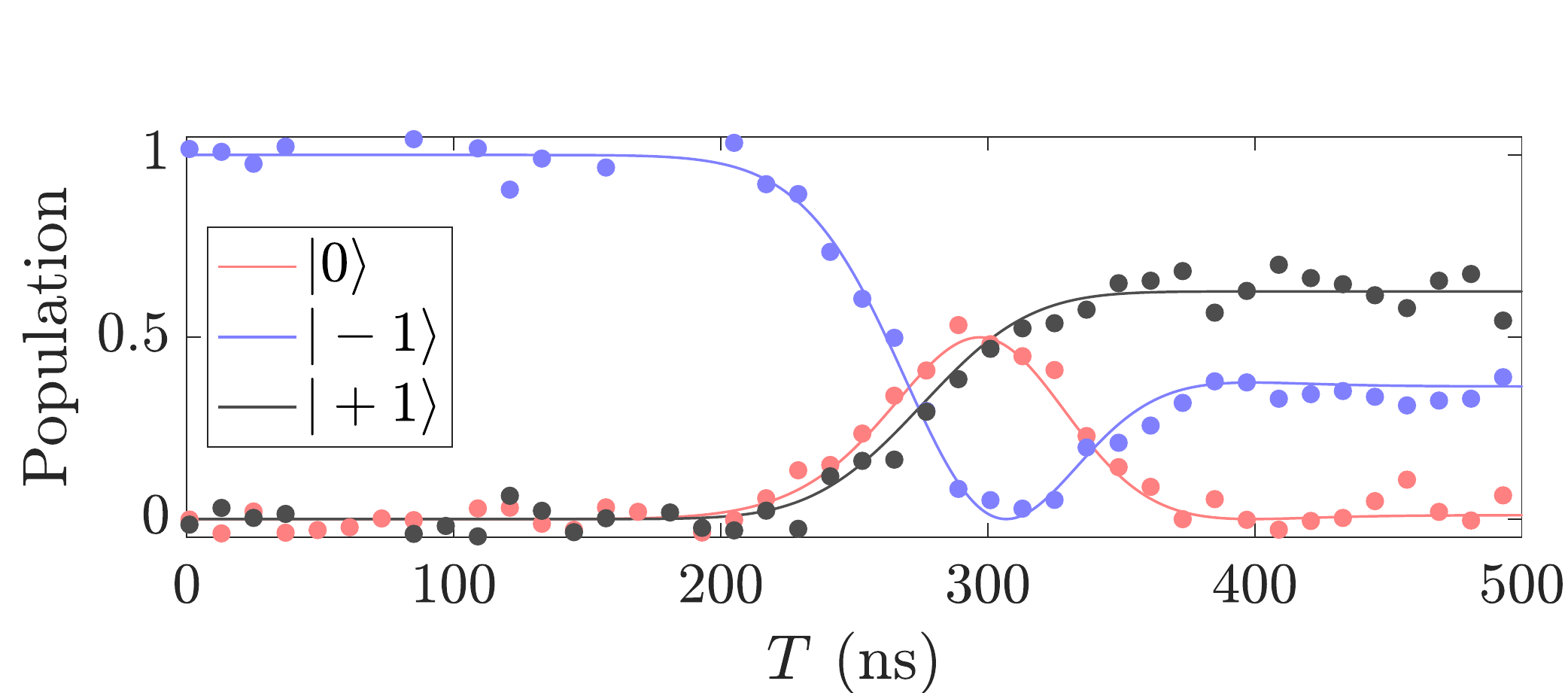}\vspace{-3ex}    \flushleft\normalsize{(b)}\hspace{-2.4ex}\includegraphics[width=\linewidth]{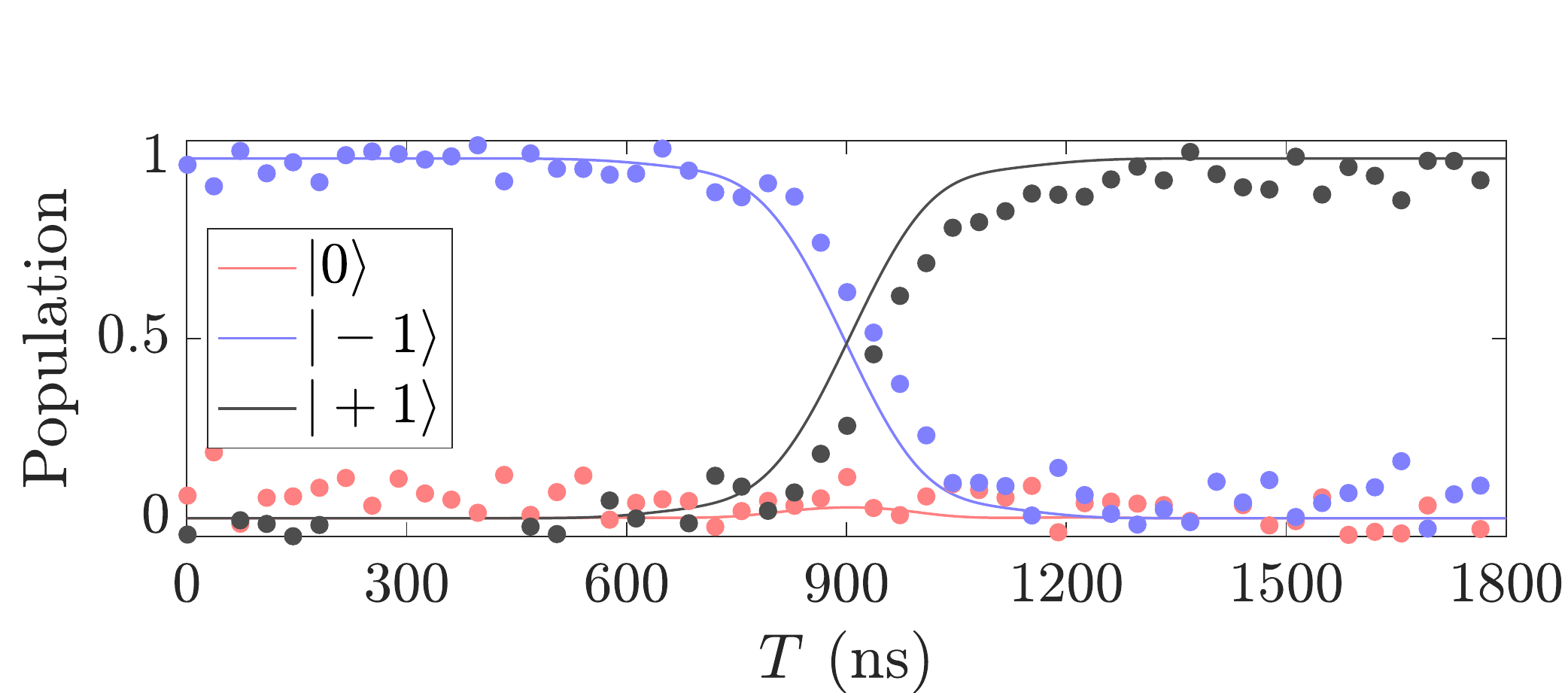}
	\caption{Evolution of the populations of the three-level system in the STIRAP protocol. (a) and (b) respectively correspond to $T=500$~ns and $T=1800$~ns. The maximum of the two control-pulse envelops $\Omega_S/2\pi$ and $\Omega_P/2\pi$ is both $8$ MHz. Markers show experimental results and solid lines are numerical simulations.}
	\label{fig:stirap_transfer_efficiency}
\end{figure}

After the adiabatic evolution, a green laser is applied to perform a fluorescence measurement and re-initialization. Then we repeat the pulse sequence adding a second $\pi$-pulse before the laser pulse [see Fig.~1(c) of the main text]. Combining the two fluorescence measurement results, one can obtain the population of all three states and thereby the state transfer efficiency. The full dynamics of the populations throughout the STIRAP process are shown in Fig.~\ref{fig:stirap_transfer_efficiency} for the two evolution times $T=500$ ns and $T=1800$ ns. We find that $>95 \%$ transfer efficiency can be achieved when the evolution time is longer than 900~ns and the transfer efficiency only reaches around $60\%$ for $T=500$~ns.


\begin{thebibliography}{89}%
	\makeatletter
	\providecommand \@ifxundefined [1]{%
		\@ifx{#1\undefined}
	}%
	\providecommand \@ifnum [1]{%
		\ifnum #1\expandafter \@firstoftwo
		\else \expandafter \@secondoftwo
		\fi
	}%
	\providecommand \@ifx [1]{%
		\ifx #1\expandafter \@firstoftwo
		\else \expandafter \@secondoftwo
		\fi
	}%
	\providecommand \natexlab [1]{#1}%
	\providecommand \enquote  [1]{``#1''}%
	\providecommand \bibnamefont  [1]{#1}%
	\providecommand \bibfnamefont [1]{#1}%
	\providecommand \citenamefont [1]{#1}%
	\providecommand \href@noop [0]{\@secondoftwo}%
	\providecommand \href [0]{\begingroup \@sanitize@url \@href}%
	\providecommand \@href[1]{\@@startlink{#1}\@@href}%
	\providecommand \@@href[1]{\endgroup#1\@@endlink}%
	\providecommand \@sanitize@url [0]{\catcode `\\12\catcode `\$12\catcode
		`\&12\catcode `\#12\catcode `\^12\catcode `\_12\catcode `\%12\relax}%
	\providecommand \@@startlink[1]{}%
	\providecommand \@@endlink[0]{}%
	\providecommand \url  [0]{\begingroup\@sanitize@url \@url }%
	\providecommand \@url [1]{\endgroup\@href {#1}{\urlprefix }}%
	\providecommand \urlprefix  [0]{URL }%
	\providecommand \Eprint [0]{\href }%
	\providecommand \doibase [0]{https://doi.org/}%
	\providecommand \selectlanguage [0]{\@gobble}%
	\providecommand \bibinfo  [0]{\@secondoftwo}%
	\providecommand \bibfield  [0]{\@secondoftwo}%
	\providecommand \translation [1]{[#1]}%
	\providecommand \BibitemOpen [0]{}%
	\providecommand \bibitemStop [0]{}%
	\providecommand \bibitemNoStop [0]{.\EOS\space}%
	\providecommand \EOS [0]{\spacefactor3000\relax}%
	\providecommand \BibitemShut  [1]{\csname bibitem#1\endcsname}%
	\let\auto@bib@innerbib\@empty
	\bibitem [{\citenamefont {Kr\'al}\ \emph {et~al.}(2007)\citenamefont {Kr\'al},
		\citenamefont {Thanopulos},\ and\ \citenamefont
		{Shapiro}}]{kral2007colloquium}%
	\BibitemOpen
	\bibfield  {author} {\bibinfo {author} {\bibfnamefont {P.}~\bibnamefont
			{Kr\'al}}, \bibinfo {author} {\bibfnamefont {I.}~\bibnamefont {Thanopulos}},\
		and\ \bibinfo {author} {\bibfnamefont {M.}~\bibnamefont {Shapiro}},\
	}\bibfield  {title} {\bibinfo {title} {Colloquium: Coherently controlled
			adiabatic passage},\ }\href {https://doi.org/10.1103/RevModPhys.79.53}
	{\bibfield  {journal} {\bibinfo  {journal} {Rev. Mod. Phys.}\ }\textbf
		{\bibinfo {volume} {79}},\ \bibinfo {pages} {53} (\bibinfo {year}
		{2007})}\BibitemShut {NoStop}%
	\bibitem [{\citenamefont {Bassett}\ \emph {et~al.}(2014)\citenamefont
		{Bassett}, \citenamefont {Heremans}, \citenamefont {Christle}, \citenamefont
		{Yale}, \citenamefont {Burkard}, \citenamefont {Buckley},\ and\ \citenamefont
		{Awschalom}}]{bassett2014ultrafast}%
	\BibitemOpen
	\bibfield  {author} {\bibinfo {author} {\bibfnamefont {L.~C.}\ \bibnamefont
			{Bassett}}, \bibinfo {author} {\bibfnamefont {F.~J.}\ \bibnamefont
			{Heremans}}, \bibinfo {author} {\bibfnamefont {D.~J.}\ \bibnamefont
			{Christle}}, \bibinfo {author} {\bibfnamefont {C.~G.}\ \bibnamefont {Yale}},
		\bibinfo {author} {\bibfnamefont {G.}~\bibnamefont {Burkard}}, \bibinfo
		{author} {\bibfnamefont {B.~B.}\ \bibnamefont {Buckley}},\ and\ \bibinfo
		{author} {\bibfnamefont {D.~D.}\ \bibnamefont {Awschalom}},\ }\bibfield
	{title} {\bibinfo {title} {Ultrafast optical control of orbital and spin
			dynamics in a solid-state defect},\ }\href
	{https://doi.org/10.1126/science.1255541} {\bibfield  {journal} {\bibinfo
			{journal} {Science}\ }\textbf {\bibinfo {volume} {345}},\ \bibinfo {pages}
		{1333} (\bibinfo {year} {2014})}\BibitemShut {NoStop}%
	\bibitem [{\citenamefont {Farhi}\ \emph {et~al.}(2001)\citenamefont {Farhi},
		\citenamefont {Goldstone}, \citenamefont {Gutmann}, \citenamefont {Lapan},
		\citenamefont {Lundgren},\ and\ \citenamefont {Preda}}]{farhi2001quantum}%
	\BibitemOpen
	\bibfield  {author} {\bibinfo {author} {\bibfnamefont {E.}~\bibnamefont
			{Farhi}}, \bibinfo {author} {\bibfnamefont {J.}~\bibnamefont {Goldstone}},
		\bibinfo {author} {\bibfnamefont {S.}~\bibnamefont {Gutmann}}, \bibinfo
		{author} {\bibfnamefont {J.}~\bibnamefont {Lapan}}, \bibinfo {author}
		{\bibfnamefont {A.}~\bibnamefont {Lundgren}},\ and\ \bibinfo {author}
		{\bibfnamefont {D.}~\bibnamefont {Preda}},\ }\bibfield  {title} {\bibinfo
		{title} {A quantum adiabatic evolution algorithm applied to random instances
			of an {NP}-complete problem},\ }\href
	{https://doi.org/10.1126/science.1057726} {\bibfield  {journal} {\bibinfo
			{journal} {Science}\ }\textbf {\bibinfo {volume} {292}},\ \bibinfo {pages}
		{472} (\bibinfo {year} {2001})}\BibitemShut {NoStop}%
	\bibitem [{\citenamefont {Monroe}\ and\ \citenamefont
		{Kim}(2013)}]{monroe2013scaling}%
	\BibitemOpen
	\bibfield  {author} {\bibinfo {author} {\bibfnamefont {C.}~\bibnamefont
			{Monroe}}\ and\ \bibinfo {author} {\bibfnamefont {J.}~\bibnamefont {Kim}},\
	}\bibfield  {title} {\bibinfo {title} {Scaling the ion trap quantum
			processor},\ }\href {https://doi.org/10.1126/science.1231298} {\bibfield
		{journal} {\bibinfo  {journal} {Science}\ }\textbf {\bibinfo {volume}
			{339}},\ \bibinfo {pages} {1164} (\bibinfo {year} {2013})}\BibitemShut
	{NoStop}%
	\bibitem [{\citenamefont {Chen}\ \emph {et~al.}(2020)\citenamefont {Chen},
		\citenamefont {Raha}, \citenamefont {Phenicie}, \citenamefont {Ourari},\ and\
		\citenamefont {Thompson}}]{chen2020parallel}%
	\BibitemOpen
	\bibfield  {author} {\bibinfo {author} {\bibfnamefont {S.}~\bibnamefont
			{Chen}}, \bibinfo {author} {\bibfnamefont {M.}~\bibnamefont {Raha}}, \bibinfo
		{author} {\bibfnamefont {C.~M.}\ \bibnamefont {Phenicie}}, \bibinfo {author}
		{\bibfnamefont {S.}~\bibnamefont {Ourari}},\ and\ \bibinfo {author}
		{\bibfnamefont {J.~D.}\ \bibnamefont {Thompson}},\ }\bibfield  {title}
	{\bibinfo {title} {Parallel single-shot measurement and coherent control of
			solid-state spins below the diffraction limit},\ }\href
	{https://doi.org/10.1126/science.abc7821} {\bibfield  {journal} {\bibinfo
			{journal} {Science}\ }\textbf {\bibinfo {volume} {370}},\ \bibinfo {pages}
		{592} (\bibinfo {year} {2020})}\BibitemShut {NoStop}%
	\bibitem [{\citenamefont {Kasevich}(2002)}]{kasevich2002coherence}%
	\BibitemOpen
	\bibfield  {author} {\bibinfo {author} {\bibfnamefont {M.~A.}\ \bibnamefont
			{Kasevich}},\ }\bibfield  {title} {\bibinfo {title} {Coherence with atoms},\
	}\href {https://doi.org/10.1126/science.1079430} {\bibfield  {journal}
		{\bibinfo  {journal} {Science}\ }\textbf {\bibinfo {volume} {298}},\ \bibinfo
		{pages} {1363} (\bibinfo {year} {2002})}\BibitemShut {NoStop}%
	\bibitem [{\citenamefont {Kotru}\ \emph {et~al.}(2015)\citenamefont {Kotru},
		\citenamefont {Butts}, \citenamefont {Kinast},\ and\ \citenamefont
		{Stoner}}]{kotru2015large}%
	\BibitemOpen
	\bibfield  {author} {\bibinfo {author} {\bibfnamefont {K.}~\bibnamefont
			{Kotru}}, \bibinfo {author} {\bibfnamefont {D.~L.}\ \bibnamefont {Butts}},
		\bibinfo {author} {\bibfnamefont {J.~M.}\ \bibnamefont {Kinast}},\ and\
		\bibinfo {author} {\bibfnamefont {R.~E.}\ \bibnamefont {Stoner}},\ }\bibfield
	{title} {\bibinfo {title} {{Large-Area Atom Interferometry with
				Frequency-Swept Raman Adiabatic Passage}},\ }\href
	{https://doi.org/10.1103/PhysRevLett.115.103001} {\bibfield  {journal}
		{\bibinfo  {journal} {Phys. Rev. Lett.}\ }\textbf {\bibinfo {volume} {115}},\
		\bibinfo {pages} {103001} (\bibinfo {year} {2015})}\BibitemShut {NoStop}%
	\bibitem [{\citenamefont {Buckley}\ \emph {et~al.}(2010)\citenamefont
		{Buckley}, \citenamefont {Fuchs}, \citenamefont {Bassett},\ and\
		\citenamefont {Awschalom}}]{buckley2010spin}%
	\BibitemOpen
	\bibfield  {author} {\bibinfo {author} {\bibfnamefont {B.}~\bibnamefont
			{Buckley}}, \bibinfo {author} {\bibfnamefont {G.}~\bibnamefont {Fuchs}},
		\bibinfo {author} {\bibfnamefont {L.}~\bibnamefont {Bassett}},\ and\ \bibinfo
		{author} {\bibfnamefont {D.}~\bibnamefont {Awschalom}},\ }\bibfield  {title}
	{\bibinfo {title} {Spin-light coherence for single-spin measurement and
			control in diamond},\ }\href {https://doi.org/10.1126/science.1196436}
	{\bibfield  {journal} {\bibinfo  {journal} {Science}\ }\textbf {\bibinfo
			{volume} {330}},\ \bibinfo {pages} {1212} (\bibinfo {year}
		{2010})}\BibitemShut {NoStop}%
	\bibitem [{\citenamefont {Liu}\ \emph {et~al.}(2020{\natexlab{a}})\citenamefont
		{Liu}, \citenamefont {Liu}, \citenamefont {Wang}, \citenamefont {Li},
		\citenamefont {Lv}, \citenamefont {Zhang}, \citenamefont {Zhang},
		\citenamefont {Teng}, \citenamefont {Zheng}, \citenamefont {Li},
		\citenamefont {Zhang}, \citenamefont {Xu},\ and\ \citenamefont
		{Gong}}]{liu2020super}%
	\BibitemOpen
	\bibfield  {author} {\bibinfo {author} {\bibfnamefont {C.}~\bibnamefont
			{Liu}}, \bibinfo {author} {\bibfnamefont {W.}~\bibnamefont {Liu}}, \bibinfo
		{author} {\bibfnamefont {S.}~\bibnamefont {Wang}}, \bibinfo {author}
		{\bibfnamefont {H.}~\bibnamefont {Li}}, \bibinfo {author} {\bibfnamefont
			{Z.}~\bibnamefont {Lv}}, \bibinfo {author} {\bibfnamefont {F.}~\bibnamefont
			{Zhang}}, \bibinfo {author} {\bibfnamefont {D.}~\bibnamefont {Zhang}},
		\bibinfo {author} {\bibfnamefont {J.}~\bibnamefont {Teng}}, \bibinfo {author}
		{\bibfnamefont {T.}~\bibnamefont {Zheng}}, \bibinfo {author} {\bibfnamefont
			{D.}~\bibnamefont {Li}}, \bibinfo {author} {\bibfnamefont {M.}~\bibnamefont
			{Zhang}}, \bibinfo {author} {\bibfnamefont {P.}~\bibnamefont {Xu}},\ and\
		\bibinfo {author} {\bibfnamefont {Q.}~\bibnamefont {Gong}},\ }\bibfield
	{title} {\bibinfo {title} {Super-resolution nanoscopy by coherent control on
			nanoparticle emission},\ }\href {https://doi.org/doi/10.1126/sciadv.aaw6579}
	{\bibfield  {journal} {\bibinfo  {journal} {Sci. Adv.}\ }\textbf {\bibinfo
			{volume} {6}},\ \bibinfo {pages} {eaaw6579} (\bibinfo {year}
		{2020}{\natexlab{a}})}\BibitemShut {NoStop}%
	\bibitem [{\citenamefont {Shapiro}\ and\ \citenamefont
		{Brumer}(2000)}]{shapiro2000coherent}%
	\BibitemOpen
	\bibfield  {author} {\bibinfo {author} {\bibfnamefont {M.}~\bibnamefont
			{Shapiro}}\ and\ \bibinfo {author} {\bibfnamefont {P.}~\bibnamefont
			{Brumer}},\ }\bibfield  {title} {\bibinfo {title} {{Coherent Control of
				Atomic, Molecular, and Electronic Processes}},\ }\href
	{https://doi.org/10.1016/S1049-250X(08)60189-5} {\bibfield  {journal}
		{\bibinfo  {journal} {Adv. At. Mol. Opt. Phys.}\ }\textbf {\bibinfo {volume}
			{42}},\ \bibinfo {pages} {287} (\bibinfo {year} {2000})}\BibitemShut
	{NoStop}%
	\bibitem [{\citenamefont {Rangelov}\ \emph {et~al.}(2005)\citenamefont
		{Rangelov}, \citenamefont {Vitanov}, \citenamefont {Yatsenko}, \citenamefont
		{Shore}, \citenamefont {Halfmann},\ and\ \citenamefont
		{Bergmann}}]{rangelov2005stark}%
	\BibitemOpen
	\bibfield  {author} {\bibinfo {author} {\bibfnamefont {A.~A.}\ \bibnamefont
			{Rangelov}}, \bibinfo {author} {\bibfnamefont {N.~V.}\ \bibnamefont
			{Vitanov}}, \bibinfo {author} {\bibfnamefont {L.~P.}\ \bibnamefont
			{Yatsenko}}, \bibinfo {author} {\bibfnamefont {B.~W.}\ \bibnamefont {Shore}},
		\bibinfo {author} {\bibfnamefont {T.}~\bibnamefont {Halfmann}},\ and\
		\bibinfo {author} {\bibfnamefont {K.}~\bibnamefont {Bergmann}},\ }\bibfield
	{title} {\bibinfo {title} {Stark-shift-chirped rapid-adiabatic-passage
			technique among three states},\ }\href
	{https://doi.org/10.1103/PhysRevA.72.053403} {\bibfield  {journal} {\bibinfo
			{journal} {Phys. Rev. A}\ }\textbf {\bibinfo {volume} {72}},\ \bibinfo
		{pages} {053403} (\bibinfo {year} {2005})}\BibitemShut {NoStop}%
	\bibitem [{\citenamefont {Torosov}\ \emph {et~al.}(2011)\citenamefont
		{Torosov}, \citenamefont {Gu\'erin},\ and\ \citenamefont
		{Vitanov}}]{torosov2011high}%
	\BibitemOpen
	\bibfield  {author} {\bibinfo {author} {\bibfnamefont {B.~T.}\ \bibnamefont
			{Torosov}}, \bibinfo {author} {\bibfnamefont {S.}~\bibnamefont {Gu\'erin}},\
		and\ \bibinfo {author} {\bibfnamefont {N.~V.}\ \bibnamefont {Vitanov}},\
	}\bibfield  {title} {\bibinfo {title} {{High-Fidelity Adiabatic Passage by
				Composite Sequences of Chirped Pulses}},\ }\href
	{https://doi.org/10.1103/PhysRevLett.106.233001} {\bibfield  {journal}
		{\bibinfo  {journal} {Phys. Rev. Lett.}\ }\textbf {\bibinfo {volume} {106}},\
		\bibinfo {pages} {233001} (\bibinfo {year} {2011})}\BibitemShut {NoStop}%
	\bibitem [{\citenamefont {Kovachy}\ \emph {et~al.}(2012)\citenamefont
		{Kovachy}, \citenamefont {Chiow},\ and\ \citenamefont
		{Kasevich}}]{kovachy2012adiabatic}%
	\BibitemOpen
	\bibfield  {author} {\bibinfo {author} {\bibfnamefont {T.}~\bibnamefont
			{Kovachy}}, \bibinfo {author} {\bibfnamefont {S.-w.}\ \bibnamefont {Chiow}},\
		and\ \bibinfo {author} {\bibfnamefont {M.~A.}\ \bibnamefont {Kasevich}},\
	}\bibfield  {title} {\bibinfo {title} {Adiabatic-rapid-passage multiphoton
			{B}ragg atom optics},\ }\href {https://doi.org/10.1103/PhysRevA.86.011606}
	{\bibfield  {journal} {\bibinfo  {journal} {Phys. Rev. A}\ }\textbf {\bibinfo
			{volume} {86}},\ \bibinfo {pages} {011606} (\bibinfo {year}
		{2012})}\BibitemShut {NoStop}%
	\bibitem [{\citenamefont {Bergmann}\ \emph {et~al.}(1998)\citenamefont
		{Bergmann}, \citenamefont {Theuer},\ and\ \citenamefont
		{Shore}}]{bergmann1998coherent}%
	\BibitemOpen
	\bibfield  {author} {\bibinfo {author} {\bibfnamefont {K.}~\bibnamefont
			{Bergmann}}, \bibinfo {author} {\bibfnamefont {H.}~\bibnamefont {Theuer}},\
		and\ \bibinfo {author} {\bibfnamefont {B.~W.}\ \bibnamefont {Shore}},\
	}\bibfield  {title} {\bibinfo {title} {Coherent population transfer among
			quantum states of atoms and molecules},\ }\href
	{https://doi.org/10.1103/RevModPhys.70.1003} {\bibfield  {journal} {\bibinfo
			{journal} {Rev. Mod. Phys.}\ }\textbf {\bibinfo {volume} {70}},\ \bibinfo
		{pages} {1003} (\bibinfo {year} {1998})}\BibitemShut {NoStop}%
	\bibitem [{\citenamefont {Fubini}\ \emph {et~al.}(2007)\citenamefont {Fubini},
		\citenamefont {Falci},\ and\ \citenamefont
		{Osterloh}}]{fubini2007robustness}%
	\BibitemOpen
	\bibfield  {author} {\bibinfo {author} {\bibfnamefont {A.}~\bibnamefont
			{Fubini}}, \bibinfo {author} {\bibfnamefont {G.}~\bibnamefont {Falci}},\ and\
		\bibinfo {author} {\bibfnamefont {A.}~\bibnamefont {Osterloh}},\ }\bibfield
	{title} {\bibinfo {title} {Robustness of adiabatic passage through a quantum
			phase transition},\ }\href {https://doi.org/10.1088/1367-2630/9/5/134}
	{\bibfield  {journal} {\bibinfo  {journal} {New J. Phys.}\ }\textbf {\bibinfo
			{volume} {9}},\ \bibinfo {pages} {134} (\bibinfo {year} {2007})}\BibitemShut
	{NoStop}%
	\bibitem [{\citenamefont {Kumar}\ \emph {et~al.}(2016)\citenamefont {Kumar},
		\citenamefont {Veps{\"a}l{\"a}inen}, \citenamefont {Danilin},\ and\
		\citenamefont {Paraoanu}}]{kumar2016stimulated}%
	\BibitemOpen
	\bibfield  {author} {\bibinfo {author} {\bibfnamefont {K.}~\bibnamefont
			{Kumar}}, \bibinfo {author} {\bibfnamefont {A.}~\bibnamefont
			{Veps{\"a}l{\"a}inen}}, \bibinfo {author} {\bibfnamefont {S.}~\bibnamefont
			{Danilin}},\ and\ \bibinfo {author} {\bibfnamefont {G.}~\bibnamefont
			{Paraoanu}},\ }\bibfield  {title} {\bibinfo {title} {Stimulated {R}aman
			adiabatic passage in a three-level superconducting circuit},\ }\href
	{https://doi.org/10.1038/ncomms10628} {\bibfield  {journal} {\bibinfo
			{journal} {Nat. Commun.}\ }\textbf {\bibinfo {volume} {7}},\ \bibinfo {pages}
		{10628} (\bibinfo {year} {2016})}\BibitemShut {NoStop}%
	\bibitem [{\citenamefont {Du}\ \emph {et~al.}(2016)\citenamefont {Du},
		\citenamefont {Liang}, \citenamefont {Li}, \citenamefont {Yue}, \citenamefont
		{Lv}, \citenamefont {Huang}, \citenamefont {Chen}, \citenamefont {Yan},\ and\
		\citenamefont {Zhu}}]{du2016experimental}%
	\BibitemOpen
	\bibfield  {author} {\bibinfo {author} {\bibfnamefont {Y.-X.}\ \bibnamefont
			{Du}}, \bibinfo {author} {\bibfnamefont {Z.-T.}\ \bibnamefont {Liang}},
		\bibinfo {author} {\bibfnamefont {Y.-C.}\ \bibnamefont {Li}}, \bibinfo
		{author} {\bibfnamefont {X.-X.}\ \bibnamefont {Yue}}, \bibinfo {author}
		{\bibfnamefont {Q.-X.}\ \bibnamefont {Lv}}, \bibinfo {author} {\bibfnamefont
			{W.}~\bibnamefont {Huang}}, \bibinfo {author} {\bibfnamefont
			{X.}~\bibnamefont {Chen}}, \bibinfo {author} {\bibfnamefont {H.}~\bibnamefont
			{Yan}},\ and\ \bibinfo {author} {\bibfnamefont {S.-L.}\ \bibnamefont {Zhu}},\
	}\bibfield  {title} {\bibinfo {title} {Experimental realization of stimulated
			{R}aman shortcut-to-adiabatic passage with cold atoms},\ }\href
	{https://doi.org/10.1038/ncomms12479} {\bibfield  {journal} {\bibinfo
			{journal} {Nat. Commun.}\ }\textbf {\bibinfo {volume} {7}},\ \bibinfo {pages}
		{12479} (\bibinfo {year} {2016})}\BibitemShut {NoStop}%
	\bibitem [{\citenamefont {Kandel}\ \emph {et~al.}(2021)\citenamefont {Kandel},
		\citenamefont {Qiao}, \citenamefont {Fallahi}, \citenamefont {Gardner},
		\citenamefont {Manfra},\ and\ \citenamefont {Nichol}}]{kandel2021adiabatic}%
	\BibitemOpen
	\bibfield  {author} {\bibinfo {author} {\bibfnamefont {Y.~P.}\ \bibnamefont
			{Kandel}}, \bibinfo {author} {\bibfnamefont {H.}~\bibnamefont {Qiao}},
		\bibinfo {author} {\bibfnamefont {S.}~\bibnamefont {Fallahi}}, \bibinfo
		{author} {\bibfnamefont {G.~C.}\ \bibnamefont {Gardner}}, \bibinfo {author}
		{\bibfnamefont {M.~J.}\ \bibnamefont {Manfra}},\ and\ \bibinfo {author}
		{\bibfnamefont {J.~M.}\ \bibnamefont {Nichol}},\ }\bibfield  {title}
	{\bibinfo {title} {Adiabatic quantum state transfer in a semiconductor
			quantum-dot spin chain},\ }\href {https://doi.org/10.1038/s41467-021-22416-5}
	{\bibfield  {journal} {\bibinfo  {journal} {Nat. Commun.}\ }\textbf {\bibinfo
			{volume} {12}},\ \bibinfo {pages} {1} (\bibinfo {year} {2021})}\BibitemShut
	{NoStop}%
	\bibitem [{\citenamefont {Vitanov}\ \emph {et~al.}(2001)\citenamefont
		{Vitanov}, \citenamefont {Halfmann}, \citenamefont {Shore},\ and\
		\citenamefont {Bergmann}}]{vitanov2001laser}%
	\BibitemOpen
	\bibfield  {author} {\bibinfo {author} {\bibfnamefont {N.~V.}\ \bibnamefont
			{Vitanov}}, \bibinfo {author} {\bibfnamefont {T.}~\bibnamefont {Halfmann}},
		\bibinfo {author} {\bibfnamefont {B.~W.}\ \bibnamefont {Shore}},\ and\
		\bibinfo {author} {\bibfnamefont {K.}~\bibnamefont {Bergmann}},\ }\bibfield
	{title} {\bibinfo {title} {Laser-induced population transfer by adiabatic
			passage techniques},\ }\href
	{https://doi.org/10.1146/annurev.physchem.52.1.763} {\bibfield  {journal}
		{\bibinfo  {journal} {Annu. Rev. Phys. Chem.}\ }\textbf {\bibinfo {volume}
			{52}},\ \bibinfo {pages} {763} (\bibinfo {year} {2001})}\BibitemShut
	{NoStop}%
	\bibitem [{\citenamefont {Aspuru-Guzik}\ \emph {et~al.}(2005)\citenamefont
		{Aspuru-Guzik}, \citenamefont {Dutoi}, \citenamefont {Love},\ and\
		\citenamefont {Head-Gordon}}]{aspuru2005simulated}%
	\BibitemOpen
	\bibfield  {author} {\bibinfo {author} {\bibfnamefont {A.}~\bibnamefont
			{Aspuru-Guzik}}, \bibinfo {author} {\bibfnamefont {A.~D.}\ \bibnamefont
			{Dutoi}}, \bibinfo {author} {\bibfnamefont {P.~J.}\ \bibnamefont {Love}},\
		and\ \bibinfo {author} {\bibfnamefont {M.}~\bibnamefont {Head-Gordon}},\
	}\bibfield  {title} {\bibinfo {title} {Simulated quantum computation of
			molecular energies},\ }\href {https://doi.org/10.1126/science.1113479}
	{\bibfield  {journal} {\bibinfo  {journal} {Science}\ }\textbf {\bibinfo
			{volume} {309}},\ \bibinfo {pages} {1704} (\bibinfo {year}
		{2005})}\BibitemShut {NoStop}%
	\bibitem [{\citenamefont {Kim}\ \emph {et~al.}(2010)\citenamefont {Kim},
		\citenamefont {Chang}, \citenamefont {Korenblit}, \citenamefont {Islam},
		\citenamefont {Edwards}, \citenamefont {Freericks}, \citenamefont {Lin},
		\citenamefont {Duan},\ and\ \citenamefont {Monroe}}]{kim2010quantum}%
	\BibitemOpen
	\bibfield  {author} {\bibinfo {author} {\bibfnamefont {K.}~\bibnamefont
			{Kim}}, \bibinfo {author} {\bibfnamefont {M.-S.}\ \bibnamefont {Chang}},
		\bibinfo {author} {\bibfnamefont {S.}~\bibnamefont {Korenblit}}, \bibinfo
		{author} {\bibfnamefont {R.}~\bibnamefont {Islam}}, \bibinfo {author}
		{\bibfnamefont {E.~E.}\ \bibnamefont {Edwards}}, \bibinfo {author}
		{\bibfnamefont {J.~K.}\ \bibnamefont {Freericks}}, \bibinfo {author}
		{\bibfnamefont {G.-D.}\ \bibnamefont {Lin}}, \bibinfo {author} {\bibfnamefont
			{L.-M.}\ \bibnamefont {Duan}},\ and\ \bibinfo {author} {\bibfnamefont
			{C.}~\bibnamefont {Monroe}},\ }\bibfield  {title} {\bibinfo {title} {Quantum
			simulation of frustrated ising spins with trapped ions},\ }\href
	{https://doi.org/10.1038/nature09071} {\bibfield  {journal} {\bibinfo
			{journal} {Nature (London)}\ }\textbf {\bibinfo {volume} {465}},\ \bibinfo
		{pages} {590} (\bibinfo {year} {2010})}\BibitemShut {NoStop}%
	\bibitem [{\citenamefont {Biamonte}\ \emph {et~al.}(2011)\citenamefont
		{Biamonte}, \citenamefont {Bergholm}, \citenamefont {Whitfield},
		\citenamefont {Fitzsimons},\ and\ \citenamefont
		{Aspuru-Guzik}}]{biamonte2011adiabatic}%
	\BibitemOpen
	\bibfield  {author} {\bibinfo {author} {\bibfnamefont {J.~D.}\ \bibnamefont
			{Biamonte}}, \bibinfo {author} {\bibfnamefont {V.}~\bibnamefont {Bergholm}},
		\bibinfo {author} {\bibfnamefont {J.~D.}\ \bibnamefont {Whitfield}}, \bibinfo
		{author} {\bibfnamefont {J.}~\bibnamefont {Fitzsimons}},\ and\ \bibinfo
		{author} {\bibfnamefont {A.}~\bibnamefont {Aspuru-Guzik}},\ }\bibfield
	{title} {\bibinfo {title} {Adiabatic quantum simulators},\ }\href
	{https://doi.org/10.1063/1.3598408} {\bibfield  {journal} {\bibinfo
			{journal} {AIP Adv.}\ }\textbf {\bibinfo {volume} {1}},\ \bibinfo {pages}
		{022126} (\bibinfo {year} {2011})}\BibitemShut {NoStop}%
	\bibitem [{\citenamefont {Jones}\ \emph {et~al.}(2000)\citenamefont {Jones},
		\citenamefont {Vedral}, \citenamefont {Ekert},\ and\ \citenamefont
		{Castagnoli}}]{jones2000geometric}%
	\BibitemOpen
	\bibfield  {author} {\bibinfo {author} {\bibfnamefont {J.~A.}\ \bibnamefont
			{Jones}}, \bibinfo {author} {\bibfnamefont {V.}~\bibnamefont {Vedral}},
		\bibinfo {author} {\bibfnamefont {A.}~\bibnamefont {Ekert}},\ and\ \bibinfo
		{author} {\bibfnamefont {G.}~\bibnamefont {Castagnoli}},\ }\bibfield  {title}
	{\bibinfo {title} {Geometric quantum computation using nuclear magnetic
			resonance},\ }\href {https://doi.org/10.1038/35002528} {\bibfield  {journal}
		{\bibinfo  {journal} {Nature (London)}\ }\textbf {\bibinfo {volume} {403}},\
		\bibinfo {pages} {869} (\bibinfo {year} {2000})}\BibitemShut {NoStop}%
	\bibitem [{\citenamefont {Barends}\ \emph {et~al.}(2016)\citenamefont
		{Barends}, \citenamefont {Shabani}, \citenamefont {Lamata}, \citenamefont
		{Kelly}, \citenamefont {Mezzacapo}, \citenamefont {Heras}, \citenamefont
		{Babbush}, \citenamefont {Fowler}, \citenamefont {Campbell}, \citenamefont
		{Chen}, \citenamefont {Chen}, \citenamefont {Chiaro}, \citenamefont
		{Dunsworth}, \citenamefont {Jeffrey}, \citenamefont {Lucero}, \citenamefont
		{Megrant}, \citenamefont {Mutus}, \citenamefont {Neeley}, \citenamefont
		{Neill}, \citenamefont {O’Malley}, \citenamefont {Quintana}, \citenamefont
		{Roushan}, \citenamefont {Sank}, \citenamefont {Vainsencher}, \citenamefont
		{Wenner}, \citenamefont {White}, \citenamefont {Solano},\ and\ \citenamefont
		{Martinis}}]{barends2016digitized}%
	\BibitemOpen
	\bibfield  {author} {\bibinfo {author} {\bibfnamefont {R.}~\bibnamefont
			{Barends}}, \bibinfo {author} {\bibfnamefont {A.}~\bibnamefont {Shabani}},
		\bibinfo {author} {\bibfnamefont {L.}~\bibnamefont {Lamata}}, \bibinfo
		{author} {\bibfnamefont {J.}~\bibnamefont {Kelly}}, \bibinfo {author}
		{\bibfnamefont {A.}~\bibnamefont {Mezzacapo}}, \bibinfo {author}
		{\bibfnamefont {U.~L.}\ \bibnamefont {Heras}}, \bibinfo {author}
		{\bibfnamefont {R.}~\bibnamefont {Babbush}}, \bibinfo {author} {\bibfnamefont
			{A.~G.}\ \bibnamefont {Fowler}}, \bibinfo {author} {\bibfnamefont
			{B.}~\bibnamefont {Campbell}}, \bibinfo {author} {\bibfnamefont
			{Y.}~\bibnamefont {Chen}}, \bibinfo {author} {\bibfnamefont {Z.}~\bibnamefont
			{Chen}}, \bibinfo {author} {\bibfnamefont {B.}~\bibnamefont {Chiaro}},
		\bibinfo {author} {\bibfnamefont {A.}~\bibnamefont {Dunsworth}}, \bibinfo
		{author} {\bibfnamefont {E.}~\bibnamefont {Jeffrey}}, \bibinfo {author}
		{\bibfnamefont {E.}~\bibnamefont {Lucero}}, \bibinfo {author} {\bibfnamefont
			{A.}~\bibnamefont {Megrant}}, \bibinfo {author} {\bibfnamefont {J.~Y.}\
			\bibnamefont {Mutus}}, \bibinfo {author} {\bibfnamefont {M.}~\bibnamefont
			{Neeley}}, \bibinfo {author} {\bibfnamefont {C.}~\bibnamefont {Neill}},
		\bibinfo {author} {\bibfnamefont {P.~J.~J.}\ \bibnamefont {O’Malley}},
		\bibinfo {author} {\bibfnamefont {C.}~\bibnamefont {Quintana}}, \bibinfo
		{author} {\bibfnamefont {P.}~\bibnamefont {Roushan}}, \bibinfo {author}
		{\bibfnamefont {D.}~\bibnamefont {Sank}}, \bibinfo {author} {\bibfnamefont
			{A.}~\bibnamefont {Vainsencher}}, \bibinfo {author} {\bibfnamefont
			{J.}~\bibnamefont {Wenner}}, \bibinfo {author} {\bibfnamefont {T.~C.}\
			\bibnamefont {White}}, \bibinfo {author} {\bibfnamefont {H.}~\bibnamefont
			{Solano}, \bibfnamefont {E.~Neven}},\ and\ \bibinfo {author} {\bibfnamefont
			{J.~M.}\ \bibnamefont {Martinis}},\ }\bibfield  {title} {\bibinfo {title}
		{Digitized adiabatic quantum computing with a superconducting circuit},\
	}\href {https://doi.org/10.1038/nature17658} {\bibfield  {journal} {\bibinfo
			{journal} {Nature (London)}\ }\textbf {\bibinfo {volume} {534}},\ \bibinfo
		{pages} {222} (\bibinfo {year} {2016})}\BibitemShut {NoStop}%
	\bibitem [{\citenamefont {Gaubatz}\ \emph {et~al.}(1990)\citenamefont
		{Gaubatz}, \citenamefont {Rudecki}, \citenamefont {Schiemann},\ and\
		\citenamefont {Bergmann}}]{gaubatz1990population}%
	\BibitemOpen
	\bibfield  {author} {\bibinfo {author} {\bibfnamefont {U.}~\bibnamefont
			{Gaubatz}}, \bibinfo {author} {\bibfnamefont {P.}~\bibnamefont {Rudecki}},
		\bibinfo {author} {\bibfnamefont {S.}~\bibnamefont {Schiemann}},\ and\
		\bibinfo {author} {\bibfnamefont {K.}~\bibnamefont {Bergmann}},\ }\bibfield
	{title} {\bibinfo {title} {Population transfer between molecular vibrational
			levels by stimulated {R}aman scattering with partially overlapping laser
			fields. {A} new concept and experimental results},\ }\href
	{https://doi.org/10.1063/1.458514} {\bibfield  {journal} {\bibinfo  {journal}
			{J. Chem. Phys.}\ }\textbf {\bibinfo {volume} {92}},\ \bibinfo {pages} {5363}
		(\bibinfo {year} {1990})}\BibitemShut {NoStop}%
	\bibitem [{\citenamefont {Vitanov}\ \emph {et~al.}(2017)\citenamefont
		{Vitanov}, \citenamefont {Rangelov}, \citenamefont {Shore},\ and\
		\citenamefont {Bergmann}}]{vitanov2017stimulated}%
	\BibitemOpen
	\bibfield  {author} {\bibinfo {author} {\bibfnamefont {N.~V.}\ \bibnamefont
			{Vitanov}}, \bibinfo {author} {\bibfnamefont {A.~A.}\ \bibnamefont
			{Rangelov}}, \bibinfo {author} {\bibfnamefont {B.~W.}\ \bibnamefont
			{Shore}},\ and\ \bibinfo {author} {\bibfnamefont {K.}~\bibnamefont
			{Bergmann}},\ }\bibfield  {title} {\bibinfo {title} {Stimulated {R}aman
			adiabatic passage in physics, chemistry, and beyond},\ }\href
	{https://doi.org/10.1103/RevModPhys.89.015006} {\bibfield  {journal}
		{\bibinfo  {journal} {Rev. Mod. Phys.}\ }\textbf {\bibinfo {volume} {89}},\
		\bibinfo {pages} {015006} (\bibinfo {year} {2017})}\BibitemShut {NoStop}%
	\bibitem [{\citenamefont {Kis}\ and\ \citenamefont
		{Renzoni}(2002)}]{kis2002qubit}%
	\BibitemOpen
	\bibfield  {author} {\bibinfo {author} {\bibfnamefont {Z.}~\bibnamefont
			{Kis}}\ and\ \bibinfo {author} {\bibfnamefont {F.}~\bibnamefont {Renzoni}},\
	}\bibfield  {title} {\bibinfo {title} {Qubit rotation by stimulated {R}aman
			adiabatic passage},\ }\href {https://doi.org/10.1103/PhysRevA.65.032318}
	{\bibfield  {journal} {\bibinfo  {journal} {Phys. Rev. A}\ }\textbf {\bibinfo
			{volume} {65}},\ \bibinfo {pages} {032318} (\bibinfo {year}
		{2002})}\BibitemShut {NoStop}%
	\bibitem [{\citenamefont {Goto}\ and\ \citenamefont
		{Ichimura}(2004)}]{goto2004multiqubit}%
	\BibitemOpen
	\bibfield  {author} {\bibinfo {author} {\bibfnamefont {H.}~\bibnamefont
			{Goto}}\ and\ \bibinfo {author} {\bibfnamefont {K.}~\bibnamefont
			{Ichimura}},\ }\bibfield  {title} {\bibinfo {title} {Multiqubit controlled
			unitary gate by adiabatic passage with an optical cavity},\ }\href
	{https://doi.org/10.1103/PhysRevA.70.012305} {\bibfield  {journal} {\bibinfo
			{journal} {Phys. Rev. A}\ }\textbf {\bibinfo {volume} {70}},\ \bibinfo
		{pages} {012305} (\bibinfo {year} {2004})}\BibitemShut {NoStop}%
	\bibitem [{\citenamefont {Zheng}(2005)}]{zheng2005nongeometric}%
	\BibitemOpen
	\bibfield  {author} {\bibinfo {author} {\bibfnamefont {S.-B.}\ \bibnamefont
			{Zheng}},\ }\bibfield  {title} {\bibinfo {title} {{Nongeometric Conditional
				Phase Shift via Adiabatic Evolution of Dark Eigenstates: A New Approach to
				Quantum Computation}},\ }\href
	{https://doi.org/10.1103/PhysRevLett.95.080502} {\bibfield  {journal}
		{\bibinfo  {journal} {Phys. Rev. Lett.}\ }\textbf {\bibinfo {volume} {95}},\
		\bibinfo {pages} {080502} (\bibinfo {year} {2005})}\BibitemShut {NoStop}%
	\bibitem [{\citenamefont {Lacour}\ \emph {et~al.}(2006)\citenamefont {Lacour},
		\citenamefont {Sangouard}, \citenamefont {Gu\'erin},\ and\ \citenamefont
		{Jauslin}}]{lacour2006arbitrary}%
	\BibitemOpen
	\bibfield  {author} {\bibinfo {author} {\bibfnamefont {X.}~\bibnamefont
			{Lacour}}, \bibinfo {author} {\bibfnamefont {N.}~\bibnamefont {Sangouard}},
		\bibinfo {author} {\bibfnamefont {S.}~\bibnamefont {Gu\'erin}},\ and\
		\bibinfo {author} {\bibfnamefont {H.~R.}\ \bibnamefont {Jauslin}},\
	}\bibfield  {title} {\bibinfo {title} {Arbitrary state controlled-unitary
			gate by adiabatic passage},\ }\href
	{https://doi.org/10.1103/PhysRevA.73.042321} {\bibfield  {journal} {\bibinfo
			{journal} {Phys. Rev. A}\ }\textbf {\bibinfo {volume} {73}},\ \bibinfo
		{pages} {042321} (\bibinfo {year} {2006})}\BibitemShut {NoStop}%
	\bibitem [{\citenamefont {Messiah}(2014)}]{messiah2014quantum}%
	\BibitemOpen
	\bibfield  {author} {\bibinfo {author} {\bibfnamefont {A.}~\bibnamefont
			{Messiah}},\ }\href@noop {} {\emph {\bibinfo {title} {Quantum Mechanics}}}\
	(\bibinfo  {publisher} {Courier Corporation},\ \bibinfo {address} {North
		Chelmsford},\ \bibinfo {year} {2014})\BibitemShut {NoStop}%
	\bibitem [{\citenamefont {Chen}\ \emph {et~al.}(2010)\citenamefont {Chen},
		\citenamefont {Lizuain}, \citenamefont {Ruschhaupt}, \citenamefont
		{Gu\'ery-Odelin},\ and\ \citenamefont {Muga}}]{chen2010shortcut}%
	\BibitemOpen
	\bibfield  {author} {\bibinfo {author} {\bibfnamefont {X.}~\bibnamefont
			{Chen}}, \bibinfo {author} {\bibfnamefont {I.}~\bibnamefont {Lizuain}},
		\bibinfo {author} {\bibfnamefont {A.}~\bibnamefont {Ruschhaupt}}, \bibinfo
		{author} {\bibfnamefont {D.}~\bibnamefont {Gu\'ery-Odelin}},\ and\ \bibinfo
		{author} {\bibfnamefont {J.~G.}\ \bibnamefont {Muga}},\ }\bibfield  {title}
	{\bibinfo {title} {{Shortcut to Adiabatic Passage in Two- and Three-Level
				Atoms}},\ }\href {https://doi.org/10.1103/PhysRevLett.105.123003} {\bibfield
		{journal} {\bibinfo  {journal} {Phys. Rev. Lett.}\ }\textbf {\bibinfo
			{volume} {105}},\ \bibinfo {pages} {123003} (\bibinfo {year}
		{2010})}\BibitemShut {NoStop}%
	\bibitem [{\citenamefont {Giannelli}\ and\ \citenamefont
		{Arimondo}(2014)}]{giannelli2014superadiabatic}%
	\BibitemOpen
	\bibfield  {author} {\bibinfo {author} {\bibfnamefont {L.}~\bibnamefont
			{Giannelli}}\ and\ \bibinfo {author} {\bibfnamefont {E.}~\bibnamefont
			{Arimondo}},\ }\bibfield  {title} {\bibinfo {title} {Three-level
			superadiabatic quantum driving},\ }\href
	{https://doi.org/10.1103/PhysRevA.89.033419} {\bibfield  {journal} {\bibinfo
			{journal} {Phys. Rev. A}\ }\textbf {\bibinfo {volume} {89}},\ \bibinfo
		{pages} {033419} (\bibinfo {year} {2014})}\BibitemShut {NoStop}%
	\bibitem [{\citenamefont {Liang}\ \emph {et~al.}(2016)\citenamefont {Liang},
		\citenamefont {Yue}, \citenamefont {Lv}, \citenamefont {Du}, \citenamefont
		{Huang}, \citenamefont {Yan},\ and\ \citenamefont
		{Zhu}}]{liang2016superadiabatic}%
	\BibitemOpen
	\bibfield  {author} {\bibinfo {author} {\bibfnamefont {Z.-T.}\ \bibnamefont
			{Liang}}, \bibinfo {author} {\bibfnamefont {X.}~\bibnamefont {Yue}}, \bibinfo
		{author} {\bibfnamefont {Q.}~\bibnamefont {Lv}}, \bibinfo {author}
		{\bibfnamefont {Y.-X.}\ \bibnamefont {Du}}, \bibinfo {author} {\bibfnamefont
			{W.}~\bibnamefont {Huang}}, \bibinfo {author} {\bibfnamefont
			{H.}~\bibnamefont {Yan}},\ and\ \bibinfo {author} {\bibfnamefont {S.-L.}\
			\bibnamefont {Zhu}},\ }\bibfield  {title} {\bibinfo {title} {Proposal for
			implementing universal superadiabatic geometric quantum gates in
			nitrogen-vacancy centers},\ }\href
	{https://doi.org/10.1103/PhysRevA.93.040305} {\bibfield  {journal} {\bibinfo
			{journal} {Phys. Rev. A}\ }\textbf {\bibinfo {volume} {93}},\ \bibinfo
		{pages} {040305} (\bibinfo {year} {2016})}\BibitemShut {NoStop}%
	\bibitem [{\citenamefont {Baksic}\ \emph {et~al.}(2016)\citenamefont {Baksic},
		\citenamefont {Ribeiro},\ and\ \citenamefont {Clerk}}]{baksic2016speeding}%
	\BibitemOpen
	\bibfield  {author} {\bibinfo {author} {\bibfnamefont {A.}~\bibnamefont
			{Baksic}}, \bibinfo {author} {\bibfnamefont {H.}~\bibnamefont {Ribeiro}},\
		and\ \bibinfo {author} {\bibfnamefont {A.~A.}\ \bibnamefont {Clerk}},\
	}\bibfield  {title} {\bibinfo {title} {{Speeding up Adiabatic Quantum State
				Transfer by Using Dressed States}},\ }\href
	{https://doi.org/10.1103/PhysRevLett.116.230503} {\bibfield  {journal}
		{\bibinfo  {journal} {Phys. Rev. Lett.}\ }\textbf {\bibinfo {volume} {116}},\
		\bibinfo {pages} {230503} (\bibinfo {year} {2016})}\BibitemShut {NoStop}%
	\bibitem [{\citenamefont {Petiziol}\ \emph {et~al.}(2020)\citenamefont
		{Petiziol}, \citenamefont {Arimondo}, \citenamefont {Giannelli},
		\citenamefont {Mintert},\ and\ \citenamefont
		{Wimberger}}]{petiziol2020superadiabatic}%
	\BibitemOpen
	\bibfield  {author} {\bibinfo {author} {\bibfnamefont {F.}~\bibnamefont
			{Petiziol}}, \bibinfo {author} {\bibfnamefont {E.}~\bibnamefont {Arimondo}},
		\bibinfo {author} {\bibfnamefont {L.}~\bibnamefont {Giannelli}}, \bibinfo
		{author} {\bibfnamefont {F.}~\bibnamefont {Mintert}},\ and\ \bibinfo {author}
		{\bibfnamefont {S.}~\bibnamefont {Wimberger}},\ }\bibfield  {title} {\bibinfo
		{title} {Optimized three-level quantum transfers based on frequency-modulated
			optical excitations},\ }\href {https://doi.org/10.1038/s41598-020-59046-8}
	{\bibfield  {journal} {\bibinfo  {journal} {Sci. Rep.}\ }\textbf {\bibinfo
			{volume} {10}},\ \bibinfo {pages} {2185} (\bibinfo {year}
		{2020})}\BibitemShut {NoStop}%
	\bibitem [{\citenamefont {Stefanatos}\ and\ \citenamefont
		{Paspalakis}(2020)}]{stefanatos2020speeding}%
	\BibitemOpen
	\bibfield  {author} {\bibinfo {author} {\bibfnamefont {D.}~\bibnamefont
			{Stefanatos}}\ and\ \bibinfo {author} {\bibfnamefont {E.}~\bibnamefont
			{Paspalakis}},\ }\bibfield  {title} {\bibinfo {title} {Speeding up adiabatic
			passage with an optimal modified {R}oland--{C}erf protocol},\ }\href
	{https://doi.org/10.1088/1751-8121/ab7423} {\bibfield  {journal} {\bibinfo
			{journal} {J. Phys. A}\ }\textbf {\bibinfo {volume} {53}},\ \bibinfo {pages}
		{115304} (\bibinfo {year} {2020})}\BibitemShut {NoStop}%
	\bibitem [{\citenamefont {del Campo}\ and\ \citenamefont
		{Kim}(2019)}]{delcampo2019focus}%
	\BibitemOpen
	\bibfield  {author} {\bibinfo {author} {\bibfnamefont {A.}~\bibnamefont {del
				Campo}}\ and\ \bibinfo {author} {\bibfnamefont {K.}~\bibnamefont {Kim}},\
	}\bibfield  {title} {\bibinfo {title} {Focus on shortcuts to adiabaticity},\
	}\href {https://doi.org/10.1088/1367-2630/ab1437} {\bibfield  {journal}
		{\bibinfo  {journal} {New J. Phys.}\ }\textbf {\bibinfo {volume} {21}},\
		\bibinfo {pages} {050201} (\bibinfo {year} {2019})}\BibitemShut {NoStop}%
	\bibitem [{\citenamefont {Gu\'ery-Odelin}\ \emph {et~al.}(2019)\citenamefont
		{Gu\'ery-Odelin}, \citenamefont {Ruschhaupt}, \citenamefont {Kiely},
		\citenamefont {Torrontegui}, \citenamefont {Mart\'{\i}nez-Garaot},\ and\
		\citenamefont {Muga}}]{gueryodelin2019review}%
	\BibitemOpen
	\bibfield  {author} {\bibinfo {author} {\bibfnamefont {D.}~\bibnamefont
			{Gu\'ery-Odelin}}, \bibinfo {author} {\bibfnamefont {A.}~\bibnamefont
			{Ruschhaupt}}, \bibinfo {author} {\bibfnamefont {A.}~\bibnamefont {Kiely}},
		\bibinfo {author} {\bibfnamefont {E.}~\bibnamefont {Torrontegui}}, \bibinfo
		{author} {\bibfnamefont {S.}~\bibnamefont {Mart\'{\i}nez-Garaot}},\ and\
		\bibinfo {author} {\bibfnamefont {J.~G.}\ \bibnamefont {Muga}},\ }\bibfield
	{title} {\bibinfo {title} {Shortcuts to adiabaticity: Concepts, methods, and
			applications},\ }\href {https://doi.org/10.1103/RevModPhys.91.045001}
	{\bibfield  {journal} {\bibinfo  {journal} {Rev. Mod. Phys.}\ }\textbf
		{\bibinfo {volume} {91}},\ \bibinfo {pages} {045001} (\bibinfo {year}
		{2019})}\BibitemShut {NoStop}%
	\bibitem [{\citenamefont {Schaff}\ \emph {et~al.}(2011)\citenamefont {Schaff},
		\citenamefont {Capuzzi}, \citenamefont {Labeyrie},\ and\ \citenamefont
		{Vignolo}}]{schaff2011coldgases}%
	\BibitemOpen
	\bibfield  {author} {\bibinfo {author} {\bibfnamefont {J.-F.}\ \bibnamefont
			{Schaff}}, \bibinfo {author} {\bibfnamefont {P.}~\bibnamefont {Capuzzi}},
		\bibinfo {author} {\bibfnamefont {G.}~\bibnamefont {Labeyrie}},\ and\
		\bibinfo {author} {\bibfnamefont {P.}~\bibnamefont {Vignolo}},\ }\bibfield
	{title} {\bibinfo {title} {Shortcuts to adiabaticity for trapped ultracold
			gases},\ }\href {https://doi.org/10.1088/1367-2630/13/11/113017} {\bibfield
		{journal} {\bibinfo  {journal} {New J. Phys.}\ }\textbf {\bibinfo {volume}
			{13}},\ \bibinfo {pages} {113017} (\bibinfo {year} {2011})}\BibitemShut
	{NoStop}%
	\bibitem [{\citenamefont {Bason}\ \emph {et~al.}(2012)\citenamefont {Bason},
		\citenamefont {Viteau}, \citenamefont {Malossi}, \citenamefont {Huillery},
		\citenamefont {Arimondo}, \citenamefont {Ciampini}, \citenamefont {Fazio},
		\citenamefont {Giovannetti}, \citenamefont {Mannella},\ and\ \citenamefont
		{Morsch}}]{bason2012boseeinstein}%
	\BibitemOpen
	\bibfield  {author} {\bibinfo {author} {\bibfnamefont {M.~G.}\ \bibnamefont
			{Bason}}, \bibinfo {author} {\bibfnamefont {M.}~\bibnamefont {Viteau}},
		\bibinfo {author} {\bibfnamefont {N.}~\bibnamefont {Malossi}}, \bibinfo
		{author} {\bibfnamefont {P.}~\bibnamefont {Huillery}}, \bibinfo {author}
		{\bibfnamefont {E.}~\bibnamefont {Arimondo}}, \bibinfo {author}
		{\bibfnamefont {D.}~\bibnamefont {Ciampini}}, \bibinfo {author}
		{\bibfnamefont {R.}~\bibnamefont {Fazio}}, \bibinfo {author} {\bibfnamefont
			{V.}~\bibnamefont {Giovannetti}}, \bibinfo {author} {\bibfnamefont
			{R.}~\bibnamefont {Mannella}},\ and\ \bibinfo {author} {\bibfnamefont
			{O.}~\bibnamefont {Morsch}},\ }\bibfield  {title} {\bibinfo {title}
		{High-fidelity quantum driving},\ }\href {https://doi.org/10.1038/nphys2170}
	{\bibfield  {journal} {\bibinfo  {journal} {Nat. Phys.}\ }\textbf {\bibinfo
			{volume} {8}},\ \bibinfo {pages} {147} (\bibinfo {year} {2012})}\BibitemShut
	{NoStop}%
	\bibitem [{\citenamefont {Malossi}\ \emph {et~al.}(2013)\citenamefont
		{Malossi}, \citenamefont {Bason}, \citenamefont {Viteau}, \citenamefont
		{Arimondo}, \citenamefont {Mannella}, \citenamefont {Morsch},\ and\
		\citenamefont {Ciampini}}]{malossi2013boseeinstein}%
	\BibitemOpen
	\bibfield  {author} {\bibinfo {author} {\bibfnamefont {N.}~\bibnamefont
			{Malossi}}, \bibinfo {author} {\bibfnamefont {M.~G.}\ \bibnamefont {Bason}},
		\bibinfo {author} {\bibfnamefont {M.}~\bibnamefont {Viteau}}, \bibinfo
		{author} {\bibfnamefont {E.}~\bibnamefont {Arimondo}}, \bibinfo {author}
		{\bibfnamefont {R.}~\bibnamefont {Mannella}}, \bibinfo {author}
		{\bibfnamefont {O.}~\bibnamefont {Morsch}},\ and\ \bibinfo {author}
		{\bibfnamefont {D.}~\bibnamefont {Ciampini}},\ }\bibfield  {title} {\bibinfo
		{title} {{Quantum driving protocols for a two-level system: From generalized
				Landau-Zener sweeps to transitionless control}},\ }\href
	{https://doi.org/10.1103/PhysRevA.87.012116} {\bibfield  {journal} {\bibinfo
			{journal} {Phys. Rev. A}\ }\textbf {\bibinfo {volume} {87}},\ \bibinfo
		{pages} {012116} (\bibinfo {year} {2013})}\BibitemShut {NoStop}%
	\bibitem [{\citenamefont {Zhang}\ \emph {et~al.}(2013)\citenamefont {Zhang},
		\citenamefont {Shim}, \citenamefont {Niemeyer}, \citenamefont {Taniguchi},
		\citenamefont {Teraji}, \citenamefont {Abe}, \citenamefont {Onoda},
		\citenamefont {Yamamoto}, \citenamefont {Ohshima}, \citenamefont {Isoya},\
		and\ \citenamefont {Suter}}]{zhang2013nvexperiment}%
	\BibitemOpen
	\bibfield  {author} {\bibinfo {author} {\bibfnamefont {J.}~\bibnamefont
			{Zhang}}, \bibinfo {author} {\bibfnamefont {J.~H.}\ \bibnamefont {Shim}},
		\bibinfo {author} {\bibfnamefont {I.}~\bibnamefont {Niemeyer}}, \bibinfo
		{author} {\bibfnamefont {T.}~\bibnamefont {Taniguchi}}, \bibinfo {author}
		{\bibfnamefont {T.}~\bibnamefont {Teraji}}, \bibinfo {author} {\bibfnamefont
			{H.}~\bibnamefont {Abe}}, \bibinfo {author} {\bibfnamefont {S.}~\bibnamefont
			{Onoda}}, \bibinfo {author} {\bibfnamefont {T.}~\bibnamefont {Yamamoto}},
		\bibinfo {author} {\bibfnamefont {T.}~\bibnamefont {Ohshima}}, \bibinfo
		{author} {\bibfnamefont {J.}~\bibnamefont {Isoya}},\ and\ \bibinfo {author}
		{\bibfnamefont {D.}~\bibnamefont {Suter}},\ }\bibfield  {title} {\bibinfo
		{title} {{Experimental Implementation of Assisted Quantum Adiabatic Passage
				in a Single Spin}},\ }\href {https://doi.org/10.1103/PhysRevLett.110.240501}
	{\bibfield  {journal} {\bibinfo  {journal} {Phys. Rev. Lett.}\ }\textbf
		{\bibinfo {volume} {110}},\ \bibinfo {pages} {240501} (\bibinfo {year}
		{2013})}\BibitemShut {NoStop}%
	\bibitem [{\citenamefont {Xu}\ \emph {et~al.}(2019)\citenamefont {Xu},
		\citenamefont {Xie}, \citenamefont {Shi}, \citenamefont {Wang}, \citenamefont
		{Xu}, \citenamefont {Wang}, \citenamefont {Wang}, \citenamefont {Plenio},\
		and\ \citenamefont {Du}}]{xu2019breaking}%
	\BibitemOpen
	\bibfield  {author} {\bibinfo {author} {\bibfnamefont {K.}~\bibnamefont
			{Xu}}, \bibinfo {author} {\bibfnamefont {T.}~\bibnamefont {Xie}}, \bibinfo
		{author} {\bibfnamefont {F.}~\bibnamefont {Shi}}, \bibinfo {author}
		{\bibfnamefont {Z.-Y.}\ \bibnamefont {Wang}}, \bibinfo {author}
		{\bibfnamefont {X.}~\bibnamefont {Xu}}, \bibinfo {author} {\bibfnamefont
			{P.}~\bibnamefont {Wang}}, \bibinfo {author} {\bibfnamefont {Y.}~\bibnamefont
			{Wang}}, \bibinfo {author} {\bibfnamefont {M.~B.}\ \bibnamefont {Plenio}},\
		and\ \bibinfo {author} {\bibfnamefont {J.}~\bibnamefont {Du}},\ }\bibfield
	{title} {\bibinfo {title} {Breaking the quantum adiabatic speed limit by
			jumping along geodesics},\ }\href {https://doi.org/10.1126/sciadv.aax3800}
	{\bibfield  {journal} {\bibinfo  {journal} {Sci. Adv.}\ }\textbf {\bibinfo
			{volume} {5}},\ \bibinfo {pages} {eaax3800} (\bibinfo {year}
		{2019})}\BibitemShut {NoStop}%
	\bibitem [{\citenamefont {Li}\ \emph {et~al.}(2018)\citenamefont {Li},
		\citenamefont {Tan}, \citenamefont {Dai}, \citenamefont {Zhao}, \citenamefont
		{Yu},\ and\ \citenamefont {Yu}}]{li2018superconducting}%
	\BibitemOpen
	\bibfield  {author} {\bibinfo {author} {\bibfnamefont {M.}~\bibnamefont
			{Li}}, \bibinfo {author} {\bibfnamefont {X.}~\bibnamefont {Tan}}, \bibinfo
		{author} {\bibfnamefont {K.}~\bibnamefont {Dai}}, \bibinfo {author}
		{\bibfnamefont {P.}~\bibnamefont {Zhao}}, \bibinfo {author} {\bibfnamefont
			{H.}~\bibnamefont {Yu}},\ and\ \bibinfo {author} {\bibfnamefont
			{Y.}~\bibnamefont {Yu}},\ }\bibfield  {title} {\bibinfo {title}
		{Demonstration of superadiabatic population transfer in superconducting
			qubit},\ }\href {https://doi.org/10.1088/1674-1056/27/6/063202} {\bibfield
		{journal} {\bibinfo  {journal} {Chin. Phys. B}\ }\textbf {\bibinfo {volume}
			{27}},\ \bibinfo {pages} {063202} (\bibinfo {year} {2018})}\BibitemShut
	{NoStop}%
	\bibitem [{\citenamefont {Veps\"al\"ainen}\ \emph {et~al.}(2019)\citenamefont
		{Veps\"al\"ainen}, \citenamefont {Danilin},\ and\ \citenamefont
		{Paraoanu}}]{vepsalainen2019superconducting}%
	\BibitemOpen
	\bibfield  {author} {\bibinfo {author} {\bibfnamefont {A.}~\bibnamefont
			{Veps\"al\"ainen}}, \bibinfo {author} {\bibfnamefont {S.}~\bibnamefont
			{Danilin}},\ and\ \bibinfo {author} {\bibfnamefont {G.~S.}\ \bibnamefont
			{Paraoanu}},\ }\bibfield  {title} {\bibinfo {title} {Superadiabatic
			population transfer in a three-level superconducting circuit},\ }\href
	{https://doi.org/10.1126/sciadv.aau5999} {\bibfield  {journal} {\bibinfo
			{journal} {Sci. Adv.}\ }\textbf {\bibinfo {volume} {5}},\ \bibinfo {pages}
		{aau5999} (\bibinfo {year} {2019})}\BibitemShut {NoStop}%
	\bibitem [{\citenamefont {Zheng}\ \emph {et~al.}(2020)\citenamefont {Zheng},
		\citenamefont {Zhang}, \citenamefont {Dong}, \citenamefont {Xu},
		\citenamefont {Wang}, \citenamefont {Wang}, \citenamefont {Li}, \citenamefont
		{Lan}, \citenamefont {Zhao}, \citenamefont {Li}, \citenamefont {Tan},\ and\
		\citenamefont {Yu}}]{zheng2022superconducting}%
	\BibitemOpen
	\bibfield  {author} {\bibinfo {author} {\bibfnamefont {W.}~\bibnamefont
			{Zheng}}, \bibinfo {author} {\bibfnamefont {Y.}~\bibnamefont {Zhang}},
		\bibinfo {author} {\bibfnamefont {Y.}~\bibnamefont {Dong}}, \bibinfo {author}
		{\bibfnamefont {J.}~\bibnamefont {Xu}}, \bibinfo {author} {\bibfnamefont
			{Z.}~\bibnamefont {Wang}}, \bibinfo {author} {\bibfnamefont {X.}~\bibnamefont
			{Wang}}, \bibinfo {author} {\bibfnamefont {Y.}~\bibnamefont {Li}}, \bibinfo
		{author} {\bibfnamefont {D.}~\bibnamefont {Lan}}, \bibinfo {author}
		{\bibfnamefont {J.}~\bibnamefont {Zhao}}, \bibinfo {author} {\bibfnamefont
			{S.}~\bibnamefont {Li}}, \bibinfo {author} {\bibfnamefont {X.}~\bibnamefont
			{Tan}},\ and\ \bibinfo {author} {\bibfnamefont {Y.}~\bibnamefont {Yu}},\
	}\bibfield  {title} {\bibinfo {title} {Optimal control of stimulated {R}aman
			adiabatic passage in a superconducting qudit},\ }\href
	{https://doi.org/10.1038/s41534-022-00521-7} {\bibfield  {journal} {\bibinfo
			{journal} {npj Quantum Inf.}\ }\textbf {\bibinfo {volume} {8}},\ \bibinfo
		{pages} {9} (\bibinfo {year} {2020})}\BibitemShut {NoStop}%
	\bibitem [{\citenamefont {Zheng}\ \emph {et~al.}()\citenamefont {Zheng},
		\citenamefont {Xu}, \citenamefont {Wang}, \citenamefont {Dong}, \citenamefont
		{Lan}, \citenamefont {Tan},\ and\ \citenamefont {Yu}}]{zheng2022arxiv}%
	\BibitemOpen
	\bibfield  {author} {\bibinfo {author} {\bibfnamefont {W.}~\bibnamefont
			{Zheng}}, \bibinfo {author} {\bibfnamefont {J.}~\bibnamefont {Xu}}, \bibinfo
		{author} {\bibfnamefont {Z.}~\bibnamefont {Wang}}, \bibinfo {author}
		{\bibfnamefont {Y.}~\bibnamefont {Dong}}, \bibinfo {author} {\bibfnamefont
			{D.}~\bibnamefont {Lan}}, \bibinfo {author} {\bibfnamefont {X.}~\bibnamefont
			{Tan}},\ and\ \bibinfo {author} {\bibfnamefont {Y.}~\bibnamefont {Yu}},\
	}\bibfield  {title} {\bibinfo {title} {{Accelerated quantum adiabatic
				transfer in superconducting qubits}},\ }\href@noop {} {\bibfield  {journal}
		{\bibinfo  {journal} {Phys. Rev. Applied}\ }\textbf {\bibinfo {volume}
			{18}},\ \bibinfo {pages} {044014}}\BibitemShut {NoStop}%
	\bibitem [{\citenamefont {Wang}\ and\ \citenamefont
		{Plenio}(2016)}]{wang2016necessary}%
	\BibitemOpen
	\bibfield  {author} {\bibinfo {author} {\bibfnamefont {Z.-Y.}\ \bibnamefont
			{Wang}}\ and\ \bibinfo {author} {\bibfnamefont {M.~B.}\ \bibnamefont
			{Plenio}},\ }\bibfield  {title} {\bibinfo {title} {Necessary and sufficient
			condition for quantum adiabatic evolution by unitary control fields},\ }\href
	{https://doi.org/10.1103/PhysRevA.93.052107} {\bibfield  {journal} {\bibinfo
			{journal} {Phys. Rev. A}\ }\textbf {\bibinfo {volume} {93}},\ \bibinfo
		{pages} {052107} (\bibinfo {year} {2016})}\BibitemShut {NoStop}%
	\bibitem [{\citenamefont {Gruber}\ \emph {et~al.}(1997)\citenamefont {Gruber},
		\citenamefont {Drabenstedt}, \citenamefont {Tietz}, \citenamefont {Fleury},
		\citenamefont {Wrachtrup},\ and\ \citenamefont {von
			Borczyskowski}}]{gruber1997scanning}%
	\BibitemOpen
	\bibfield  {author} {\bibinfo {author} {\bibfnamefont {A.}~\bibnamefont
			{Gruber}}, \bibinfo {author} {\bibfnamefont {A.}~\bibnamefont {Drabenstedt}},
		\bibinfo {author} {\bibfnamefont {C.}~\bibnamefont {Tietz}}, \bibinfo
		{author} {\bibfnamefont {L.}~\bibnamefont {Fleury}}, \bibinfo {author}
		{\bibfnamefont {J.}~\bibnamefont {Wrachtrup}},\ and\ \bibinfo {author}
		{\bibfnamefont {C.}~\bibnamefont {von Borczyskowski}},\ }\bibfield  {title}
	{\bibinfo {title} {Scanning confocal optical microscopy and magnetic
			resonance on single defect centers},\ }\href
	{https://doi.org/10.1126/science.276.5321.2012} {\bibfield  {journal}
		{\bibinfo  {journal} {Science}\ }\textbf {\bibinfo {volume} {276}},\ \bibinfo
		{pages} {2012} (\bibinfo {year} {1997})}\BibitemShut {NoStop}%
	\bibitem [{\citenamefont {Rondin}\ \emph {et~al.}(2014)\citenamefont {Rondin},
		\citenamefont {Tetienne}, \citenamefont {Hingant}, \citenamefont {Roch},
		\citenamefont {Maletinsky},\ and\ \citenamefont
		{Jacques}}]{rondin2014magnetometry}%
	\BibitemOpen
	\bibfield  {author} {\bibinfo {author} {\bibfnamefont {L.}~\bibnamefont
			{Rondin}}, \bibinfo {author} {\bibfnamefont {J.-P.}\ \bibnamefont
			{Tetienne}}, \bibinfo {author} {\bibfnamefont {T.}~\bibnamefont {Hingant}},
		\bibinfo {author} {\bibfnamefont {J.-F.}\ \bibnamefont {Roch}}, \bibinfo
		{author} {\bibfnamefont {P.}~\bibnamefont {Maletinsky}},\ and\ \bibinfo
		{author} {\bibfnamefont {V.}~\bibnamefont {Jacques}},\ }\bibfield  {title}
	{\bibinfo {title} {Magnetometry with nitrogen-vacancy defects in diamond},\
	}\href {https://doi.org/10.1088/0034-4885/77/5/056503} {\bibfield  {journal}
		{\bibinfo  {journal} {Rep. Prog. Phys.}\ }\textbf {\bibinfo {volume} {77}},\
		\bibinfo {pages} {056503} (\bibinfo {year} {2014})}\BibitemShut {NoStop}%
	\bibitem [{\citenamefont {Jelezko}\ \emph {et~al.}(2004)\citenamefont
		{Jelezko}, \citenamefont {Gaebel}, \citenamefont {Popa}, \citenamefont
		{Gruber},\ and\ \citenamefont {Wrachtrup}}]{jelezko2004observation}%
	\BibitemOpen
	\bibfield  {author} {\bibinfo {author} {\bibfnamefont {F.}~\bibnamefont
			{Jelezko}}, \bibinfo {author} {\bibfnamefont {T.}~\bibnamefont {Gaebel}},
		\bibinfo {author} {\bibfnamefont {I.}~\bibnamefont {Popa}}, \bibinfo {author}
		{\bibfnamefont {A.}~\bibnamefont {Gruber}},\ and\ \bibinfo {author}
		{\bibfnamefont {J.}~\bibnamefont {Wrachtrup}},\ }\bibfield  {title} {\bibinfo
		{title} {{Observation of Coherent Oscillations in a Single Electron Spin}},\
	}\href {https://doi.org/10.1103/PhysRevLett.92.076401} {\bibfield  {journal}
		{\bibinfo  {journal} {Phys. Rev. Lett.}\ }\textbf {\bibinfo {volume} {92}},\
		\bibinfo {pages} {076401} (\bibinfo {year} {2004})}\BibitemShut {NoStop}%
	\bibitem [{\citenamefont {Bar-Gill}\ \emph {et~al.}(2013)\citenamefont
		{Bar-Gill}, \citenamefont {Pham}, \citenamefont {Jarmola}, \citenamefont
		{Budker},\ and\ \citenamefont {Walsworth}}]{bar2013solid}%
	\BibitemOpen
	\bibfield  {author} {\bibinfo {author} {\bibfnamefont {N.}~\bibnamefont
			{Bar-Gill}}, \bibinfo {author} {\bibfnamefont {L.~M.}\ \bibnamefont {Pham}},
		\bibinfo {author} {\bibfnamefont {A.}~\bibnamefont {Jarmola}}, \bibinfo
		{author} {\bibfnamefont {D.}~\bibnamefont {Budker}},\ and\ \bibinfo {author}
		{\bibfnamefont {R.~L.}\ \bibnamefont {Walsworth}},\ }\bibfield  {title}
	{\bibinfo {title} {Solid-state electronic spin coherence time approaching one
			second},\ }\href {https://doi.org/10.1038/ncomms2771} {\bibfield  {journal}
		{\bibinfo  {journal} {Nat. Commun.}\ }\textbf {\bibinfo {volume} {4}},\
		\bibinfo {pages} {1} (\bibinfo {year} {2013})}\BibitemShut {NoStop}%
	\bibitem [{\citenamefont {Balasubramanian}\ \emph {et~al.}(2009)\citenamefont
		{Balasubramanian}, \citenamefont {Neumann}, \citenamefont {Twitchen},
		\citenamefont {Markham}, \citenamefont {Kolesov}, \citenamefont {Mizuochi},
		\citenamefont {Isoya}, \citenamefont {Achard}, \citenamefont {Beck},
		\citenamefont {Tissler}, \citenamefont {Jacques}, \citenamefont {Hemmer},
		\citenamefont {Jelezko},\ and\ \citenamefont
		{Wrachtrup}}]{balasubramanian2009ultralong}%
	\BibitemOpen
	\bibfield  {author} {\bibinfo {author} {\bibfnamefont {G.}~\bibnamefont
			{Balasubramanian}}, \bibinfo {author} {\bibfnamefont {P.}~\bibnamefont
			{Neumann}}, \bibinfo {author} {\bibfnamefont {D.}~\bibnamefont {Twitchen}},
		\bibinfo {author} {\bibfnamefont {M.}~\bibnamefont {Markham}}, \bibinfo
		{author} {\bibfnamefont {R.}~\bibnamefont {Kolesov}}, \bibinfo {author}
		{\bibfnamefont {N.}~\bibnamefont {Mizuochi}}, \bibinfo {author}
		{\bibfnamefont {J.}~\bibnamefont {Isoya}}, \bibinfo {author} {\bibfnamefont
			{J.}~\bibnamefont {Achard}}, \bibinfo {author} {\bibfnamefont
			{J.}~\bibnamefont {Beck}}, \bibinfo {author} {\bibfnamefont {J.}~\bibnamefont
			{Tissler}}, \bibinfo {author} {\bibfnamefont {V.}~\bibnamefont {Jacques}},
		\bibinfo {author} {\bibfnamefont {P.~R.}\ \bibnamefont {Hemmer}}, \bibinfo
		{author} {\bibfnamefont {F.}~\bibnamefont {Jelezko}},\ and\ \bibinfo {author}
		{\bibfnamefont {J.}~\bibnamefont {Wrachtrup}},\ }\bibfield  {title} {\bibinfo
		{title} {Ultralong spin coherence time in isotopically engineered diamond},\
	}\href {https://doi.org/10.1038/nmat2420} {\bibfield  {journal} {\bibinfo
			{journal} {Nat. Mater.}\ }\textbf {\bibinfo {volume} {8}},\ \bibinfo {pages}
		{383} (\bibinfo {year} {2009})}\BibitemShut {NoStop}%
	\bibitem [{\citenamefont {Chu}\ \emph {et~al.}(2015)\citenamefont {Chu},
		\citenamefont {Markham}, \citenamefont {Twitchen},\ and\ \citenamefont
		{Lukin}}]{chu2015all}%
	\BibitemOpen
	\bibfield  {author} {\bibinfo {author} {\bibfnamefont {Y.}~\bibnamefont
			{Chu}}, \bibinfo {author} {\bibfnamefont {M.}~\bibnamefont {Markham}},
		\bibinfo {author} {\bibfnamefont {D.~J.}\ \bibnamefont {Twitchen}},\ and\
		\bibinfo {author} {\bibfnamefont {M.~D.}\ \bibnamefont {Lukin}},\ }\bibfield
	{title} {\bibinfo {title} {All-optical control of a single electron spin in
			diamond},\ }\href {https://doi.org/10.1103/PhysRevA.91.021801} {\bibfield
		{journal} {\bibinfo  {journal} {Phys. Rev. A}\ }\textbf {\bibinfo {volume}
			{91}},\ \bibinfo {pages} {021801} (\bibinfo {year} {2015})}\BibitemShut
	{NoStop}%
	\bibitem [{\citenamefont {Shu}\ \emph {et~al.}(2018)\citenamefont {Shu},
		\citenamefont {Liu}, \citenamefont {Cao}, \citenamefont {Yang}, \citenamefont
		{Zhang}, \citenamefont {Plenio}, \citenamefont {Jelezko},\ and\ \citenamefont
		{Cai}}]{shu2018observation}%
	\BibitemOpen
	\bibfield  {author} {\bibinfo {author} {\bibfnamefont {Z.}~\bibnamefont
			{Shu}}, \bibinfo {author} {\bibfnamefont {Y.}~\bibnamefont {Liu}}, \bibinfo
		{author} {\bibfnamefont {Q.}~\bibnamefont {Cao}}, \bibinfo {author}
		{\bibfnamefont {P.}~\bibnamefont {Yang}}, \bibinfo {author} {\bibfnamefont
			{S.}~\bibnamefont {Zhang}}, \bibinfo {author} {\bibfnamefont {M.~B.}\
			\bibnamefont {Plenio}}, \bibinfo {author} {\bibfnamefont {F.}~\bibnamefont
			{Jelezko}},\ and\ \bibinfo {author} {\bibfnamefont {J.}~\bibnamefont {Cai}},\
	}\bibfield  {title} {\bibinfo {title} {{Observation of Floquet Raman
				Transition in a Driven Solid-State Spin System}},\ }\href
	{https://doi.org/10.1103/PhysRevLett.121.210501} {\bibfield  {journal}
		{\bibinfo  {journal} {Phys. Rev. Lett.}\ }\textbf {\bibinfo {volume} {121}},\
		\bibinfo {pages} {210501} (\bibinfo {year} {2018})}\BibitemShut {NoStop}%
	\bibitem [{\citenamefont {Tian}\ \emph {et~al.}(2020)\citenamefont {Tian},
		\citenamefont {Liu}, \citenamefont {Liu}, \citenamefont {Yang}, \citenamefont
		{Betzholz}, \citenamefont {Said}, \citenamefont {Jelezko},\ and\
		\citenamefont {Cai}}]{tian2020quantum}%
	\BibitemOpen
	\bibfield  {author} {\bibinfo {author} {\bibfnamefont {J.}~\bibnamefont
			{Tian}}, \bibinfo {author} {\bibfnamefont {H.}~\bibnamefont {Liu}}, \bibinfo
		{author} {\bibfnamefont {Y.}~\bibnamefont {Liu}}, \bibinfo {author}
		{\bibfnamefont {P.}~\bibnamefont {Yang}}, \bibinfo {author} {\bibfnamefont
			{R.}~\bibnamefont {Betzholz}}, \bibinfo {author} {\bibfnamefont {R.~S.}\
			\bibnamefont {Said}}, \bibinfo {author} {\bibfnamefont {F.}~\bibnamefont
			{Jelezko}},\ and\ \bibinfo {author} {\bibfnamefont {J.}~\bibnamefont {Cai}},\
	}\bibfield  {title} {\bibinfo {title} {Quantum optimal control using
			phase-modulated driving fields},\ }\href
	{https://doi.org/10.1103/PhysRevA.102.043707} {\bibfield  {journal} {\bibinfo
			{journal} {Phys. Rev. A}\ }\textbf {\bibinfo {volume} {102}},\ \bibinfo
		{pages} {043707} (\bibinfo {year} {2020})}\BibitemShut {NoStop}%
	\bibitem [{\citenamefont {Cao}\ \emph {et~al.}(2020)\citenamefont {Cao},
		\citenamefont {Yang}, \citenamefont {Gong}, \citenamefont {Yu}, \citenamefont
		{Retzker}, \citenamefont {Plenio}, \citenamefont {M\"uller}, \citenamefont
		{Tomek}, \citenamefont {Naydenov}, \citenamefont {McGuinness}, \citenamefont
		{Jelezko},\ and\ \citenamefont {Cai}}]{cao2020pra}%
	\BibitemOpen
	\bibfield  {author} {\bibinfo {author} {\bibfnamefont {Q.-Y.}\ \bibnamefont
			{Cao}}, \bibinfo {author} {\bibfnamefont {P.-C.}\ \bibnamefont {Yang}},
		\bibinfo {author} {\bibfnamefont {M.-S.}\ \bibnamefont {Gong}}, \bibinfo
		{author} {\bibfnamefont {M.}~\bibnamefont {Yu}}, \bibinfo {author}
		{\bibfnamefont {A.}~\bibnamefont {Retzker}}, \bibinfo {author} {\bibfnamefont
			{M.}~\bibnamefont {Plenio}}, \bibinfo {author} {\bibfnamefont
			{C.}~\bibnamefont {M\"uller}}, \bibinfo {author} {\bibfnamefont
			{N.}~\bibnamefont {Tomek}}, \bibinfo {author} {\bibfnamefont
			{B.}~\bibnamefont {Naydenov}}, \bibinfo {author} {\bibfnamefont
			{L.}~\bibnamefont {McGuinness}}, \bibinfo {author} {\bibfnamefont
			{F.}~\bibnamefont {Jelezko}},\ and\ \bibinfo {author} {\bibfnamefont {J.-M.}\
			\bibnamefont {Cai}},\ }\bibfield  {title} {\bibinfo {title} {{Protecting
				Quantum Spin Coherence of Nanodiamonds in Living Cells}},\ }\href
	{https://doi.org/10.1103/PhysRevApplied.13.024021} {\bibfield  {journal}
		{\bibinfo  {journal} {Phys. Rev. Applied}\ }\textbf {\bibinfo {volume}
			{13}},\ \bibinfo {pages} {024021} (\bibinfo {year} {2020})}\BibitemShut
	{NoStop}%
	\bibitem [{\citenamefont {Wrachtrup}\ and\ \citenamefont
		{Jelezko}(2006)}]{wrachtrup2006processing}%
	\BibitemOpen
	\bibfield  {author} {\bibinfo {author} {\bibfnamefont {J.}~\bibnamefont
			{Wrachtrup}}\ and\ \bibinfo {author} {\bibfnamefont {F.}~\bibnamefont
			{Jelezko}},\ }\bibfield  {title} {\bibinfo {title} {Processing quantum
			information in diamond},\ }\href
	{https://doi.org/10.1088/0953-8984/18/21/S08} {\bibfield  {journal} {\bibinfo
			{journal} {J. Phys.: Condens. Matter}\ }\textbf {\bibinfo {volume} {18}},\
		\bibinfo {pages} {S807} (\bibinfo {year} {2006})}\BibitemShut {NoStop}%
	\bibitem [{\citenamefont {Dolde}\ \emph {et~al.}(2013)\citenamefont {Dolde},
		\citenamefont {Jakobi}, \citenamefont {Naydenov}, \citenamefont {Zhao},
		\citenamefont {Pezzagna}, \citenamefont {Trautmann}, \citenamefont {Meijer},
		\citenamefont {Neumann}, \citenamefont {Jelezko},\ and\ \citenamefont
		{Wrachtrup}}]{dolde2013room}%
	\BibitemOpen
	\bibfield  {author} {\bibinfo {author} {\bibfnamefont {F.}~\bibnamefont
			{Dolde}}, \bibinfo {author} {\bibfnamefont {I.}~\bibnamefont {Jakobi}},
		\bibinfo {author} {\bibfnamefont {B.}~\bibnamefont {Naydenov}}, \bibinfo
		{author} {\bibfnamefont {N.}~\bibnamefont {Zhao}}, \bibinfo {author}
		{\bibfnamefont {S.}~\bibnamefont {Pezzagna}}, \bibinfo {author}
		{\bibfnamefont {C.}~\bibnamefont {Trautmann}}, \bibinfo {author}
		{\bibfnamefont {J.}~\bibnamefont {Meijer}}, \bibinfo {author} {\bibfnamefont
			{P.}~\bibnamefont {Neumann}}, \bibinfo {author} {\bibfnamefont
			{F.}~\bibnamefont {Jelezko}},\ and\ \bibinfo {author} {\bibfnamefont
			{J.}~\bibnamefont {Wrachtrup}},\ }\bibfield  {title} {\bibinfo {title}
		{Room-temperature entanglement between single defect spins in diamond},\
	}\href {https://doi.org/10.1038/nphys2545} {\bibfield  {journal} {\bibinfo
			{journal} {Nat. Phys.}\ }\textbf {\bibinfo {volume} {9}},\ \bibinfo {pages}
		{139} (\bibinfo {year} {2013})}\BibitemShut {NoStop}%
	\bibitem [{\citenamefont {Cai}\ \emph {et~al.}(2013)\citenamefont {Cai},
		\citenamefont {Retzker}, \citenamefont {Jelezko},\ and\ \citenamefont
		{Plenio}}]{cai2013large}%
	\BibitemOpen
	\bibfield  {author} {\bibinfo {author} {\bibfnamefont {J.}~\bibnamefont
			{Cai}}, \bibinfo {author} {\bibfnamefont {A.}~\bibnamefont {Retzker}},
		\bibinfo {author} {\bibfnamefont {F.}~\bibnamefont {Jelezko}},\ and\ \bibinfo
		{author} {\bibfnamefont {M.~B.}\ \bibnamefont {Plenio}},\ }\bibfield  {title}
	{\bibinfo {title} {A large-scale quantum simulator on a diamond surface at
			room temperature},\ }\href {https://doi.org/10.1038/nphys2519} {\bibfield
		{journal} {\bibinfo  {journal} {Nat. Phys.}\ }\textbf {\bibinfo {volume}
			{9}},\ \bibinfo {pages} {168} (\bibinfo {year} {2013})}\BibitemShut {NoStop}%
	\bibitem [{\citenamefont {Xiang}\ \emph {et~al.}(2013)\citenamefont {Xiang},
		\citenamefont {Ashhab}, \citenamefont {You},\ and\ \citenamefont
		{Nori}}]{xiang2013hybrid}%
	\BibitemOpen
	\bibfield  {author} {\bibinfo {author} {\bibfnamefont {Z.-L.}\ \bibnamefont
			{Xiang}}, \bibinfo {author} {\bibfnamefont {S.}~\bibnamefont {Ashhab}},
		\bibinfo {author} {\bibfnamefont {J.~Q.}\ \bibnamefont {You}},\ and\ \bibinfo
		{author} {\bibfnamefont {F.}~\bibnamefont {Nori}},\ }\bibfield  {title}
	{\bibinfo {title} {Hybrid quantum circuits: Superconducting circuits
			interacting with other quantum systems},\ }\href
	{https://doi.org/10.1103/RevModPhys.85.623} {\bibfield  {journal} {\bibinfo
			{journal} {Rev. Mod. Phys.}\ }\textbf {\bibinfo {volume} {85}},\ \bibinfo
		{pages} {623} (\bibinfo {year} {2013})}\BibitemShut {NoStop}%
	\bibitem [{\citenamefont {Kurizki}\ \emph {et~al.}(2015)\citenamefont
		{Kurizki}, \citenamefont {Bertet}, \citenamefont {Kubo}, \citenamefont
		{M{\o}lmer}, \citenamefont {Petrosyan}, \citenamefont {Rabl},\ and\
		\citenamefont {Schmiedmayer}}]{kurizki2015quantum}%
	\BibitemOpen
	\bibfield  {author} {\bibinfo {author} {\bibfnamefont {G.}~\bibnamefont
			{Kurizki}}, \bibinfo {author} {\bibfnamefont {P.}~\bibnamefont {Bertet}},
		\bibinfo {author} {\bibfnamefont {Y.}~\bibnamefont {Kubo}}, \bibinfo {author}
		{\bibfnamefont {K.}~\bibnamefont {M{\o}lmer}}, \bibinfo {author}
		{\bibfnamefont {D.}~\bibnamefont {Petrosyan}}, \bibinfo {author}
		{\bibfnamefont {P.}~\bibnamefont {Rabl}},\ and\ \bibinfo {author}
		{\bibfnamefont {J.}~\bibnamefont {Schmiedmayer}},\ }\bibfield  {title}
	{\bibinfo {title} {Quantum technologies with hybrid systems},\ }\href
	{https://doi.org/10.1073/pnas.1419326112} {\bibfield  {journal} {\bibinfo
			{journal} {Proc. Natl. Acad. Sci. U.S.A.}\ }\textbf {\bibinfo {volume}
			{112}},\ \bibinfo {pages} {3866} (\bibinfo {year} {2015})}\BibitemShut
	{NoStop}%
	\bibitem [{\citenamefont {Yu}\ \emph {et~al.}(2020)\citenamefont {Yu},
		\citenamefont {Yang}, \citenamefont {Gong}, \citenamefont {Cao},
		\citenamefont {Lu}, \citenamefont {Liu}, \citenamefont {Zhang}, \citenamefont
		{Plenio}, \citenamefont {Jelezko}, \citenamefont {Ozawa}, \citenamefont
		{Goldman},\ and\ \citenamefont {Cai}}]{yu2020experimental}%
	\BibitemOpen
	\bibfield  {author} {\bibinfo {author} {\bibfnamefont {M.}~\bibnamefont
			{Yu}}, \bibinfo {author} {\bibfnamefont {P.}~\bibnamefont {Yang}}, \bibinfo
		{author} {\bibfnamefont {M.}~\bibnamefont {Gong}}, \bibinfo {author}
		{\bibfnamefont {Q.}~\bibnamefont {Cao}}, \bibinfo {author} {\bibfnamefont
			{Q.}~\bibnamefont {Lu}}, \bibinfo {author} {\bibfnamefont {H.}~\bibnamefont
			{Liu}}, \bibinfo {author} {\bibfnamefont {S.}~\bibnamefont {Zhang}}, \bibinfo
		{author} {\bibfnamefont {M.~B.}\ \bibnamefont {Plenio}}, \bibinfo {author}
		{\bibfnamefont {F.}~\bibnamefont {Jelezko}}, \bibinfo {author} {\bibfnamefont
			{T.}~\bibnamefont {Ozawa}}, \bibinfo {author} {\bibfnamefont
			{N.}~\bibnamefont {Goldman}},\ and\ \bibinfo {author} {\bibfnamefont
			{j.}~\bibnamefont {Cai}},\ }\bibfield  {title} {\bibinfo {title}
		{Experimental measurement of the quantum geometric tensor using coupled
			qubits in diamond},\ }\href {https://doi.org/10.1093/nsr/nwz193} {\bibfield
		{journal} {\bibinfo  {journal} {Nat. Sci. Rev.}\ }\textbf {\bibinfo {volume}
			{7}},\ \bibinfo {pages} {254} (\bibinfo {year} {2020})}\BibitemShut {NoStop}%
	\bibitem [{\citenamefont {Yu}\ \emph {et~al.}(2022)\citenamefont {Yu},
		\citenamefont {Liu}, \citenamefont {Yang}, \citenamefont {Gong},
		\citenamefont {Cao}, \citenamefont {Zhang}, \citenamefont {Liu},
		\citenamefont {Heyl}, \citenamefont {Ozawa}, \citenamefont {Goldman},\ and\
		\citenamefont {Cai}}]{yu2022quantum}%
	\BibitemOpen
	\bibfield  {author} {\bibinfo {author} {\bibfnamefont {M.}~\bibnamefont
			{Yu}}, \bibinfo {author} {\bibfnamefont {Y.}~\bibnamefont {Liu}}, \bibinfo
		{author} {\bibfnamefont {P.}~\bibnamefont {Yang}}, \bibinfo {author}
		{\bibfnamefont {M.}~\bibnamefont {Gong}}, \bibinfo {author} {\bibfnamefont
			{Q.}~\bibnamefont {Cao}}, \bibinfo {author} {\bibfnamefont {S.}~\bibnamefont
			{Zhang}}, \bibinfo {author} {\bibfnamefont {H.}~\bibnamefont {Liu}}, \bibinfo
		{author} {\bibfnamefont {M.}~\bibnamefont {Heyl}}, \bibinfo {author}
		{\bibfnamefont {T.}~\bibnamefont {Ozawa}}, \bibinfo {author} {\bibfnamefont
			{N.}~\bibnamefont {Goldman}},\ and\ \bibinfo {author} {\bibfnamefont
			{J.}~\bibnamefont {Cai}},\ }\bibfield  {title} {\bibinfo {title} {Quantum
			{F}isher information measurement and verification of the quantum
			{C}ram{\'e}r--{R}ao bound in a solid-state qubit},\ }\href
	{https://doi.org/10.1038/s41534-022-00547-x} {\bibfield  {journal} {\bibinfo
			{journal} {npj Quantum Inf.}\ }\textbf {\bibinfo {volume} {8}},\ \bibinfo
		{pages} {56} (\bibinfo {year} {2022})}\BibitemShut {NoStop}%
	\bibitem [{\citenamefont {Yu}\ \emph {et~al.}(2021)\citenamefont {Yu},
		\citenamefont {Li}, \citenamefont {Wang}, \citenamefont {Chu}, \citenamefont
		{Yang}, \citenamefont {Gong}, \citenamefont {Goldman},\ and\ \citenamefont
		{Cai}}]{yu2021experimental}%
	\BibitemOpen
	\bibfield  {author} {\bibinfo {author} {\bibfnamefont {M.}~\bibnamefont
			{Yu}}, \bibinfo {author} {\bibfnamefont {D.}~\bibnamefont {Li}}, \bibinfo
		{author} {\bibfnamefont {J.}~\bibnamefont {Wang}}, \bibinfo {author}
		{\bibfnamefont {Y.}~\bibnamefont {Chu}}, \bibinfo {author} {\bibfnamefont
			{P.}~\bibnamefont {Yang}}, \bibinfo {author} {\bibfnamefont {M.}~\bibnamefont
			{Gong}}, \bibinfo {author} {\bibfnamefont {N.}~\bibnamefont {Goldman}},\ and\
		\bibinfo {author} {\bibfnamefont {J.}~\bibnamefont {Cai}},\ }\bibfield
	{title} {\bibinfo {title} {Experimental estimation of the quantum {F}isher
			information from randomized measurements},\ }\href
	{https://doi.org/10.1103/PhysRevResearch.3.043122} {\bibfield  {journal}
		{\bibinfo  {journal} {Phys. Rev. Research}\ }\textbf {\bibinfo {volume}
			{3}},\ \bibinfo {pages} {043122} (\bibinfo {year} {2021})}\BibitemShut
	{NoStop}%
	\bibitem [{\citenamefont {Neumann}\ \emph {et~al.}(2008)\citenamefont
		{Neumann}, \citenamefont {Mizuochi}, \citenamefont {Rempp}, \citenamefont
		{Hemmer}, \citenamefont {Watanabe}, \citenamefont {Yamasaki}, \citenamefont
		{Jacques}, \citenamefont {Gaebel}, \citenamefont {Jelezko},\ and\
		\citenamefont {Wrachtrup}}]{neumann2008multipartite}%
	\BibitemOpen
	\bibfield  {author} {\bibinfo {author} {\bibfnamefont {P.}~\bibnamefont
			{Neumann}}, \bibinfo {author} {\bibfnamefont {N.}~\bibnamefont {Mizuochi}},
		\bibinfo {author} {\bibfnamefont {F.}~\bibnamefont {Rempp}}, \bibinfo
		{author} {\bibfnamefont {P.}~\bibnamefont {Hemmer}}, \bibinfo {author}
		{\bibfnamefont {H.}~\bibnamefont {Watanabe}}, \bibinfo {author}
		{\bibfnamefont {S.}~\bibnamefont {Yamasaki}}, \bibinfo {author}
		{\bibfnamefont {V.}~\bibnamefont {Jacques}}, \bibinfo {author} {\bibfnamefont
			{T.}~\bibnamefont {Gaebel}}, \bibinfo {author} {\bibfnamefont
			{F.}~\bibnamefont {Jelezko}},\ and\ \bibinfo {author} {\bibfnamefont
			{J.}~\bibnamefont {Wrachtrup}},\ }\bibfield  {title} {\bibinfo {title}
		{Multipartite entanglement among single spins in diamond},\ }\href
	{https://doi.org/10.1126/science.1157233} {\bibfield  {journal} {\bibinfo
			{journal} {Science}\ }\textbf {\bibinfo {volume} {320}},\ \bibinfo {pages}
		{1326} (\bibinfo {year} {2008})}\BibitemShut {NoStop}%
	\bibitem [{\citenamefont {Zu}\ \emph {et~al.}(2014)\citenamefont {Zu},
		\citenamefont {Wang}, \citenamefont {He}, \citenamefont {Zhang},
		\citenamefont {Dai}, \citenamefont {Wang},\ and\ \citenamefont
		{Duan}}]{zu2014experimental}%
	\BibitemOpen
	\bibfield  {author} {\bibinfo {author} {\bibfnamefont {C.}~\bibnamefont
			{Zu}}, \bibinfo {author} {\bibfnamefont {W.-B.}\ \bibnamefont {Wang}},
		\bibinfo {author} {\bibfnamefont {L.}~\bibnamefont {He}}, \bibinfo {author}
		{\bibfnamefont {W.-G.}\ \bibnamefont {Zhang}}, \bibinfo {author}
		{\bibfnamefont {C.-Y.}\ \bibnamefont {Dai}}, \bibinfo {author} {\bibfnamefont
			{F.}~\bibnamefont {Wang}},\ and\ \bibinfo {author} {\bibfnamefont {L.-M.}\
			\bibnamefont {Duan}},\ }\bibfield  {title} {\bibinfo {title} {Experimental
			realization of universal geometric quantum gates with solid-state spins},\
	}\href {https://doi.org/10.1038/nature13729} {\bibfield  {journal} {\bibinfo
			{journal} {Nature (London)}\ }\textbf {\bibinfo {volume} {514}},\ \bibinfo
		{pages} {72} (\bibinfo {year} {2014})}\BibitemShut {NoStop}%
	\bibitem [{\citenamefont {Jelezko}\ and\ \citenamefont
		{Wrachtrup}(2006)}]{jelezko2006single}%
	\BibitemOpen
	\bibfield  {author} {\bibinfo {author} {\bibfnamefont {F.}~\bibnamefont
			{Jelezko}}\ and\ \bibinfo {author} {\bibfnamefont {J.}~\bibnamefont
			{Wrachtrup}},\ }\bibfield  {title} {\bibinfo {title} {Single defect centres
			in diamond: A review},\ }\href {https://doi.org/10.1002/pssa.200671403}
	{\bibfield  {journal} {\bibinfo  {journal} {Physica Status Solidi A}\
		}\textbf {\bibinfo {volume} {203}},\ \bibinfo {pages} {3207} (\bibinfo {year}
		{2006})}\BibitemShut {NoStop}%
	\bibitem [{\citenamefont {Doherty}\ \emph {et~al.}(2013)\citenamefont
		{Doherty}, \citenamefont {Manson}, \citenamefont {Delaney}, \citenamefont
		{Jelezko}, \citenamefont {Wrachtrup},\ and\ \citenamefont
		{Hollenberg}}]{doherty2013nitrogen}%
	\BibitemOpen
	\bibfield  {author} {\bibinfo {author} {\bibfnamefont {M.~W.}\ \bibnamefont
			{Doherty}}, \bibinfo {author} {\bibfnamefont {N.~B.}\ \bibnamefont {Manson}},
		\bibinfo {author} {\bibfnamefont {P.}~\bibnamefont {Delaney}}, \bibinfo
		{author} {\bibfnamefont {F.}~\bibnamefont {Jelezko}}, \bibinfo {author}
		{\bibfnamefont {J.}~\bibnamefont {Wrachtrup}},\ and\ \bibinfo {author}
		{\bibfnamefont {L.~C.~L.}\ \bibnamefont {Hollenberg}},\ }\bibfield  {title}
	{\bibinfo {title} {The nitrogen-vacancy colour centre in diamond},\ }\href
	{https://doi.org/10.1016/j.physrep.2013.02.001} {\bibfield  {journal}
		{\bibinfo  {journal} {Phys. Rep.}\ }\textbf {\bibinfo {volume} {528}},\
		\bibinfo {pages} {1} (\bibinfo {year} {2013})}\BibitemShut {NoStop}%
	\bibitem [{\citenamefont {Kucsko}\ \emph {et~al.}(2013)\citenamefont {Kucsko},
		\citenamefont {Maurer}, \citenamefont {Yao}, \citenamefont {Kubo},
		\citenamefont {Noh}, \citenamefont {Lo}, \citenamefont {Park},\ and\
		\citenamefont {Lukin}}]{kucsko2013nanometre}%
	\BibitemOpen
	\bibfield  {author} {\bibinfo {author} {\bibfnamefont {G.}~\bibnamefont
			{Kucsko}}, \bibinfo {author} {\bibfnamefont {P.~C.}\ \bibnamefont {Maurer}},
		\bibinfo {author} {\bibfnamefont {N.~Y.}\ \bibnamefont {Yao}}, \bibinfo
		{author} {\bibfnamefont {M.}~\bibnamefont {Kubo}}, \bibinfo {author}
		{\bibfnamefont {H.~J.}\ \bibnamefont {Noh}}, \bibinfo {author} {\bibfnamefont
			{P.~K.}\ \bibnamefont {Lo}}, \bibinfo {author} {\bibfnamefont
			{H.}~\bibnamefont {Park}},\ and\ \bibinfo {author} {\bibfnamefont {M.~D.}\
			\bibnamefont {Lukin}},\ }\bibfield  {title} {\bibinfo {title}
		{Nanometre-scale thermometry in a living cell},\ }\href
	{https://doi.org/10.1038/nature12373} {\bibfield  {journal} {\bibinfo
			{journal} {Nature (London)}\ }\textbf {\bibinfo {volume} {500}},\ \bibinfo
		{pages} {54} (\bibinfo {year} {2013})}\BibitemShut {NoStop}%
	\bibitem [{\citenamefont {Acosta}\ \emph {et~al.}(2010)\citenamefont {Acosta},
		\citenamefont {Bauch}, \citenamefont {Ledbetter}, \citenamefont {Waxman},
		\citenamefont {Bouchard},\ and\ \citenamefont {Budker}}]{Acosta2010prl}%
	\BibitemOpen
	\bibfield  {author} {\bibinfo {author} {\bibfnamefont {V.~M.}\ \bibnamefont
			{Acosta}}, \bibinfo {author} {\bibfnamefont {E.}~\bibnamefont {Bauch}},
		\bibinfo {author} {\bibfnamefont {M.~P.}\ \bibnamefont {Ledbetter}}, \bibinfo
		{author} {\bibfnamefont {A.}~\bibnamefont {Waxman}}, \bibinfo {author}
		{\bibfnamefont {L.-S.}\ \bibnamefont {Bouchard}},\ and\ \bibinfo {author}
		{\bibfnamefont {D.}~\bibnamefont {Budker}},\ }\bibfield  {title} {\bibinfo
		{title} {Temperature dependence of the nitrogen-vacancy magnetic resonance in
			diamond},\ }\href {https://doi.org/10.1103/PhysRevLett.104.070801} {\bibfield
		{journal} {\bibinfo  {journal} {Phys. Rev. Lett.}\ }\textbf {\bibinfo
			{volume} {104}},\ \bibinfo {pages} {070801} (\bibinfo {year}
		{2010})}\BibitemShut {NoStop}%
	\bibitem [{\citenamefont {Neumann}\ \emph {et~al.}(2013)\citenamefont
		{Neumann}, \citenamefont {Jakobi}, \citenamefont {Dolde}, \citenamefont
		{Burk}, \citenamefont {Reuter}, \citenamefont {Waldherr}, \citenamefont
		{Honert}, \citenamefont {Wolf}, \citenamefont {Brunner}, \citenamefont
		{Shim}, \citenamefont {Suter}, \citenamefont {Sumiya}, \citenamefont
		{Isoya},\ and\ \citenamefont {Wrachtrup}}]{neumann2013nl}%
	\BibitemOpen
	\bibfield  {author} {\bibinfo {author} {\bibfnamefont {P.}~\bibnamefont
			{Neumann}}, \bibinfo {author} {\bibfnamefont {I.}~\bibnamefont {Jakobi}},
		\bibinfo {author} {\bibfnamefont {F.}~\bibnamefont {Dolde}}, \bibinfo
		{author} {\bibfnamefont {C.}~\bibnamefont {Burk}}, \bibinfo {author}
		{\bibfnamefont {R.}~\bibnamefont {Reuter}}, \bibinfo {author} {\bibfnamefont
			{G.}~\bibnamefont {Waldherr}}, \bibinfo {author} {\bibfnamefont
			{J.}~\bibnamefont {Honert}}, \bibinfo {author} {\bibfnamefont
			{T.}~\bibnamefont {Wolf}}, \bibinfo {author} {\bibfnamefont {A.}~\bibnamefont
			{Brunner}}, \bibinfo {author} {\bibfnamefont {J.~H.}\ \bibnamefont {Shim}},
		\bibinfo {author} {\bibfnamefont {D.}~\bibnamefont {Suter}}, \bibinfo
		{author} {\bibfnamefont {H.}~\bibnamefont {Sumiya}}, \bibinfo {author}
		{\bibfnamefont {J.}~\bibnamefont {Isoya}},\ and\ \bibinfo {author}
		{\bibfnamefont {J.}~\bibnamefont {Wrachtrup}},\ }\bibfield  {title} {\bibinfo
		{title} {High-precision nanoscale temperature sensing using single defects in
			diamond},\ }\href {https://doi.org/10.1021/nl401216y} {\bibfield  {journal}
		{\bibinfo  {journal} {Nano Lett.}\ }\textbf {\bibinfo {volume} {13}},\
		\bibinfo {pages} {2738} (\bibinfo {year} {2013})}\BibitemShut {NoStop}%
	\bibitem [{\citenamefont {Plakhotnik}\ \emph {et~al.}(2014)\citenamefont
		{Plakhotnik}, \citenamefont {Doherty}, \citenamefont {Cole}, \citenamefont
		{Chapman},\ and\ \citenamefont {Manson}}]{plakhotnik2014nl}%
	\BibitemOpen
	\bibfield  {author} {\bibinfo {author} {\bibfnamefont {T.}~\bibnamefont
			{Plakhotnik}}, \bibinfo {author} {\bibfnamefont {M.~W.}\ \bibnamefont
			{Doherty}}, \bibinfo {author} {\bibfnamefont {J.~H.}\ \bibnamefont {Cole}},
		\bibinfo {author} {\bibfnamefont {R.}~\bibnamefont {Chapman}},\ and\ \bibinfo
		{author} {\bibfnamefont {N.~B.}\ \bibnamefont {Manson}},\ }\bibfield  {title}
	{\bibinfo {title} {All-optical thermometry and thermal properties of the
			optically detected spin resonances of the {NV}–center in nanodiamond},\
	}\href {https://doi.org/10.1021/nl501841d} {\bibfield  {journal} {\bibinfo
			{journal} {Nano Lett.}\ }\textbf {\bibinfo {volume} {14}},\ \bibinfo {pages}
		{4989} (\bibinfo {year} {2014})}\BibitemShut {NoStop}%
	\bibitem [{\citenamefont {Knauer}\ \emph {et~al.}(2020)\citenamefont {Knauer},
		\citenamefont {Hadden},\ and\ \citenamefont {Rarity}}]{knauer2020situ}%
	\BibitemOpen
	\bibfield  {author} {\bibinfo {author} {\bibfnamefont {S.}~\bibnamefont
			{Knauer}}, \bibinfo {author} {\bibfnamefont {J.~P.}\ \bibnamefont {Hadden}},\
		and\ \bibinfo {author} {\bibfnamefont {J.~G.}\ \bibnamefont {Rarity}},\
	}\bibfield  {title} {\bibinfo {title} {In-situ measurements of fabrication
			induced strain in diamond photonic-structures using intrinsic colour
			centres},\ }\href {https://doi.org/10.1038/s41534-020-0277-1} {\bibfield
		{journal} {\bibinfo  {journal} {npj Quantum Inf.}\ }\textbf {\bibinfo
			{volume} {6}},\ \bibinfo {pages} {1} (\bibinfo {year} {2020})}\BibitemShut
	{NoStop}%
	\bibitem [{\citenamefont {{Trusheim}}\ and\ \citenamefont
		{{Englund}}(2016)}]{trusheim2016njp}%
	\BibitemOpen
	\bibfield  {author} {\bibinfo {author} {\bibfnamefont {M.~E.}\ \bibnamefont
			{{Trusheim}}}\ and\ \bibinfo {author} {\bibfnamefont {D.}~\bibnamefont
			{{Englund}}},\ }\bibfield  {title} {\bibinfo {title} {Wide-field strain
			imaging with preferentially aligned nitrogen-vacancy centers in
			polycrystalline diamond},\ }\href {https://doi.org/10.1088/1367-2630/aa5040}
	{\bibfield  {journal} {\bibinfo  {journal} {New J. Phys.}\ }\textbf {\bibinfo
			{volume} {18}},\ \bibinfo {pages} {123023} (\bibinfo {year}
		{2016})}\BibitemShut {NoStop}%
	\bibitem [{\citenamefont {Kehayias}\ \emph {et~al.}(2019)\citenamefont
		{Kehayias}, \citenamefont {Turner}, \citenamefont {Trubko}, \citenamefont
		{Schloss}, \citenamefont {Hart}, \citenamefont {Wesson}, \citenamefont
		{Glenn},\ and\ \citenamefont {Walsworth}}]{kehayias2019prb}%
	\BibitemOpen
	\bibfield  {author} {\bibinfo {author} {\bibfnamefont {P.}~\bibnamefont
			{Kehayias}}, \bibinfo {author} {\bibfnamefont {M.~J.}\ \bibnamefont
			{Turner}}, \bibinfo {author} {\bibfnamefont {R.}~\bibnamefont {Trubko}},
		\bibinfo {author} {\bibfnamefont {J.~M.}\ \bibnamefont {Schloss}}, \bibinfo
		{author} {\bibfnamefont {C.~A.}\ \bibnamefont {Hart}}, \bibinfo {author}
		{\bibfnamefont {M.}~\bibnamefont {Wesson}}, \bibinfo {author} {\bibfnamefont
			{D.~R.}\ \bibnamefont {Glenn}},\ and\ \bibinfo {author} {\bibfnamefont
			{R.~L.}\ \bibnamefont {Walsworth}},\ }\bibfield  {title} {\bibinfo {title}
		{Imaging crystal stress in diamond using ensembles of nitrogen-vacancy
			centers},\ }\href {https://doi.org/10.1103/PhysRevB.100.174103} {\bibfield
		{journal} {\bibinfo  {journal} {Phys. Rev. B}\ }\textbf {\bibinfo {volume}
			{100}},\ \bibinfo {pages} {174103} (\bibinfo {year} {2019})}\BibitemShut
	{NoStop}%
	\bibitem [{\citenamefont {Dolde}\ \emph {et~al.}(2011)\citenamefont {Dolde},
		\citenamefont {Fedder}, \citenamefont {Doherty}, \citenamefont {N{\"o}bauer},
		\citenamefont {Rempp}, \citenamefont {Balasubramanian}, \citenamefont {Wolf},
		\citenamefont {Reinhard}, \citenamefont {Hollenberg}, \citenamefont
		{Jelezko},\ and\ \citenamefont {Wrachtrup}}]{dolde2011electric}%
	\BibitemOpen
	\bibfield  {author} {\bibinfo {author} {\bibfnamefont {F.}~\bibnamefont
			{Dolde}}, \bibinfo {author} {\bibfnamefont {H.}~\bibnamefont {Fedder}},
		\bibinfo {author} {\bibfnamefont {M.~W.}\ \bibnamefont {Doherty}}, \bibinfo
		{author} {\bibfnamefont {T.}~\bibnamefont {N{\"o}bauer}}, \bibinfo {author}
		{\bibfnamefont {F.}~\bibnamefont {Rempp}}, \bibinfo {author} {\bibfnamefont
			{G.}~\bibnamefont {Balasubramanian}}, \bibinfo {author} {\bibfnamefont
			{T.}~\bibnamefont {Wolf}}, \bibinfo {author} {\bibfnamefont {F.}~\bibnamefont
			{Reinhard}}, \bibinfo {author} {\bibfnamefont {L.~C.~L.}\ \bibnamefont
			{Hollenberg}}, \bibinfo {author} {\bibfnamefont {F.}~\bibnamefont
			{Jelezko}},\ and\ \bibinfo {author} {\bibfnamefont {J.}~\bibnamefont
			{Wrachtrup}},\ }\bibfield  {title} {\bibinfo {title} {Electric-field sensing
			using single diamond spins},\ }\href {https://doi.org/10.1038/nphys1969}
	{\bibfield  {journal} {\bibinfo  {journal} {Nat. Phys.}\ }\textbf {\bibinfo
			{volume} {7}},\ \bibinfo {pages} {459} (\bibinfo {year} {2011})}\BibitemShut
	{NoStop}%
	\bibitem [{\citenamefont {Maze}\ \emph {et~al.}(2008)\citenamefont {Maze},
		\citenamefont {Stanwix}, \citenamefont {Hodges}, \citenamefont {Hong},
		\citenamefont {Taylor}, \citenamefont {Cappellaro}, \citenamefont {Jiang},
		\citenamefont {Dutt}, \citenamefont {Togan}, \citenamefont {Zibrov},
		\citenamefont {Yacoby}, \citenamefont {Walsworth},\ and\ \citenamefont
		{Lukin}}]{maze2008nanoscale}%
	\BibitemOpen
	\bibfield  {author} {\bibinfo {author} {\bibfnamefont {J.~R.}\ \bibnamefont
			{Maze}}, \bibinfo {author} {\bibfnamefont {P.~L.}\ \bibnamefont {Stanwix}},
		\bibinfo {author} {\bibfnamefont {J.~S.}\ \bibnamefont {Hodges}}, \bibinfo
		{author} {\bibfnamefont {S.}~\bibnamefont {Hong}}, \bibinfo {author}
		{\bibfnamefont {J.~M.}\ \bibnamefont {Taylor}}, \bibinfo {author}
		{\bibfnamefont {P.}~\bibnamefont {Cappellaro}}, \bibinfo {author}
		{\bibfnamefont {L.}~\bibnamefont {Jiang}}, \bibinfo {author} {\bibfnamefont
			{M.~V.}\ \bibnamefont {Dutt}}, \bibinfo {author} {\bibfnamefont
			{E.}~\bibnamefont {Togan}}, \bibinfo {author} {\bibfnamefont {A.~S.}\
			\bibnamefont {Zibrov}}, \bibinfo {author} {\bibfnamefont {A.}~\bibnamefont
			{Yacoby}}, \bibinfo {author} {\bibfnamefont {R.~L.}\ \bibnamefont
			{Walsworth}},\ and\ \bibinfo {author} {\bibfnamefont {M.~D.}\ \bibnamefont
			{Lukin}},\ }\bibfield  {title} {\bibinfo {title} {Nanoscale magnetic sensing
			with an individual electronic spin in diamond},\ }\href
	{https://doi.org/10.1038/nature07279} {\bibfield  {journal} {\bibinfo
			{journal} {Nature (London)}\ }\textbf {\bibinfo {volume} {455}},\ \bibinfo
		{pages} {644} (\bibinfo {year} {2008})}\BibitemShut {NoStop}%
	\bibitem [{\citenamefont {Balasubramanian}\ \emph {et~al.}(2008)\citenamefont
		{Balasubramanian}, \citenamefont {Chan}, \citenamefont {Kolesov},
		\citenamefont {Al-Hmoud}, \citenamefont {Tisler}, \citenamefont {Shin},
		\citenamefont {Kim}, \citenamefont {Wojcik}, \citenamefont {Hemmer},
		\citenamefont {Krueger}, \citenamefont {Hanke}, \citenamefont
		{Leitenstorfer}, \citenamefont {Bratschitsch}, \citenamefont {Jelezko},\ and\
		\citenamefont {Wrachtrup}}]{balasubramanian2008nanoscale}%
	\BibitemOpen
	\bibfield  {author} {\bibinfo {author} {\bibfnamefont {G.}~\bibnamefont
			{Balasubramanian}}, \bibinfo {author} {\bibfnamefont {I.}~\bibnamefont
			{Chan}}, \bibinfo {author} {\bibfnamefont {R.}~\bibnamefont {Kolesov}},
		\bibinfo {author} {\bibfnamefont {M.}~\bibnamefont {Al-Hmoud}}, \bibinfo
		{author} {\bibfnamefont {J.}~\bibnamefont {Tisler}}, \bibinfo {author}
		{\bibfnamefont {C.}~\bibnamefont {Shin}}, \bibinfo {author} {\bibfnamefont
			{C.}~\bibnamefont {Kim}}, \bibinfo {author} {\bibfnamefont {A.}~\bibnamefont
			{Wojcik}}, \bibinfo {author} {\bibfnamefont {P.~R.}\ \bibnamefont {Hemmer}},
		\bibinfo {author} {\bibfnamefont {A.}~\bibnamefont {Krueger}}, \bibinfo
		{author} {\bibfnamefont {T.}~\bibnamefont {Hanke}}, \bibinfo {author}
		{\bibfnamefont {A.}~\bibnamefont {Leitenstorfer}}, \bibinfo {author}
		{\bibfnamefont {R.}~\bibnamefont {Bratschitsch}}, \bibinfo {author}
		{\bibfnamefont {F.}~\bibnamefont {Jelezko}},\ and\ \bibinfo {author}
		{\bibfnamefont {J.}~\bibnamefont {Wrachtrup}},\ }\bibfield  {title} {\bibinfo
		{title} {Nanoscale imaging magnetometry with diamond spins under ambient
			conditions},\ }\href {https://doi.org/10.1038/nature07278} {\bibfield
		{journal} {\bibinfo  {journal} {Nature (London)}\ }\textbf {\bibinfo {volume}
			{455}},\ \bibinfo {pages} {648} (\bibinfo {year} {2008})}\BibitemShut
	{NoStop}%
	\bibitem [{\citenamefont {Welter}\ \emph {et~al.}(2022)\citenamefont {Welter},
		\citenamefont {Rhensius}, \citenamefont {Morales}, \citenamefont {W\"ornlei},
		\citenamefont {Lambert}, \citenamefont {Puebla-Hellmann}, \citenamefont
		{Gambardella},\ and\ \citenamefont {Degen}}]{welter2022magnetometry}%
	\BibitemOpen
	\bibfield  {author} {\bibinfo {author} {\bibfnamefont {P.}~\bibnamefont
			{Welter}}, \bibinfo {author} {\bibfnamefont {J.}~\bibnamefont {Rhensius}},
		\bibinfo {author} {\bibfnamefont {A.}~\bibnamefont {Morales}}, \bibinfo
		{author} {\bibfnamefont {M.~S.}\ \bibnamefont {W\"ornlei}}, \bibinfo {author}
		{\bibfnamefont {C.-H.}\ \bibnamefont {Lambert}}, \bibinfo {author}
		{\bibfnamefont {G.}~\bibnamefont {Puebla-Hellmann}}, \bibinfo {author}
		{\bibfnamefont {P.}~\bibnamefont {Gambardella}},\ and\ \bibinfo {author}
		{\bibfnamefont {C.~L.}\ \bibnamefont {Degen}},\ }\bibfield  {title} {\bibinfo
		{title} {Scanning nitrogen-vacancy center magnetometry in large in-plane
			magnetic fields},\ }\href {https://doi.org/10.1063/5.0084910} {\bibfield
		{journal} {\bibinfo  {journal} {Appl. Phys. Lett.}\ }\textbf {\bibinfo
			{volume} {120}},\ \bibinfo {pages} {074003} (\bibinfo {year}
		{2022})}\BibitemShut {NoStop}%
	\bibitem [{\citenamefont {Cai}\ \emph {et~al.}(2014)\citenamefont {Cai},
		\citenamefont {Jelezko},\ and\ \citenamefont {Plenio}}]{cai2014nc}%
	\BibitemOpen
	\bibfield  {author} {\bibinfo {author} {\bibfnamefont {J.}~\bibnamefont
			{Cai}}, \bibinfo {author} {\bibfnamefont {F.}~\bibnamefont {Jelezko}},\ and\
		\bibinfo {author} {\bibfnamefont {M.~B.}\ \bibnamefont {Plenio}},\ }\bibfield
	{title} {\bibinfo {title} {Hybrid sensors based on colour centres in diamond
			and piezoactive layers},\ }\href {https://doi.org/10.1038/ncomms5065}
	{\bibfield  {journal} {\bibinfo  {journal} {Nat. Commun.}\ }\textbf {\bibinfo
			{volume} {5}},\ \bibinfo {pages} {4065} (\bibinfo {year} {2014})}\BibitemShut
	{NoStop}%
	\bibitem [{\citenamefont {Wang}\ \emph {et~al.}(2018)\citenamefont {Wang},
		\citenamefont {Liu}, \citenamefont {Leong}, \citenamefont {Zeng},
		\citenamefont {Feng}, \citenamefont {Li}, \citenamefont {Dolde},
		\citenamefont {Fedder}, \citenamefont {Wrachtrup}, \citenamefont {Cui},
		\citenamefont {Yang}, \citenamefont {Li},\ and\ \citenamefont
		{Liu}}]{wang2018prx}%
	\BibitemOpen
	\bibfield  {author} {\bibinfo {author} {\bibfnamefont {N.}~\bibnamefont
			{Wang}}, \bibinfo {author} {\bibfnamefont {G.-Q.}\ \bibnamefont {Liu}},
		\bibinfo {author} {\bibfnamefont {W.-H.}\ \bibnamefont {Leong}}, \bibinfo
		{author} {\bibfnamefont {H.}~\bibnamefont {Zeng}}, \bibinfo {author}
		{\bibfnamefont {X.}~\bibnamefont {Feng}}, \bibinfo {author} {\bibfnamefont
			{S.-H.}\ \bibnamefont {Li}}, \bibinfo {author} {\bibfnamefont
			{F.}~\bibnamefont {Dolde}}, \bibinfo {author} {\bibfnamefont
			{H.}~\bibnamefont {Fedder}}, \bibinfo {author} {\bibfnamefont
			{J.}~\bibnamefont {Wrachtrup}}, \bibinfo {author} {\bibfnamefont {X.-D.}\
			\bibnamefont {Cui}}, \bibinfo {author} {\bibfnamefont {S.}~\bibnamefont
			{Yang}}, \bibinfo {author} {\bibfnamefont {Q.}~\bibnamefont {Li}},\ and\
		\bibinfo {author} {\bibfnamefont {R.-B.}\ \bibnamefont {Liu}},\ }\bibfield
	{title} {\bibinfo {title} {{Magnetic Criticality Enhanced Hybrid Nanodiamond
				Thermometer under Ambient Conditions}},\ }\href
	{https://doi.org/10.1103/PhysRevX.8.011042} {\bibfield  {journal} {\bibinfo
			{journal} {Phys. Rev. X}\ }\textbf {\bibinfo {volume} {8}},\ \bibinfo {pages}
		{011042} (\bibinfo {year} {2018})}\BibitemShut {NoStop}%
	\bibitem [{\citenamefont {Liu}\ \emph {et~al.}(2020{\natexlab{b}})\citenamefont
		{Liu}, \citenamefont {Leong}, \citenamefont {Xia}, \citenamefont {Feng},
		\citenamefont {Finkler}, \citenamefont {Denisenko}, \citenamefont
		{Wrachtrup}, \citenamefont {Li},\ and\ \citenamefont {Liu}}]{liu2020nsr}%
	\BibitemOpen
	\bibfield  {author} {\bibinfo {author} {\bibfnamefont {C.-F.}\ \bibnamefont
			{Liu}}, \bibinfo {author} {\bibfnamefont {W.-H.}\ \bibnamefont {Leong}},
		\bibinfo {author} {\bibfnamefont {K.}~\bibnamefont {Xia}}, \bibinfo {author}
		{\bibfnamefont {X.}~\bibnamefont {Feng}}, \bibinfo {author} {\bibfnamefont
			{A.}~\bibnamefont {Finkler}}, \bibinfo {author} {\bibfnamefont
			{A.}~\bibnamefont {Denisenko}}, \bibinfo {author} {\bibfnamefont
			{J.}~\bibnamefont {Wrachtrup}}, \bibinfo {author} {\bibfnamefont
			{Q.}~\bibnamefont {Li}},\ and\ \bibinfo {author} {\bibfnamefont {R.-B.}\
			\bibnamefont {Liu}},\ }\bibfield  {title} {\bibinfo {title} {Ultra-sensitive
			hybrid diamond nanothermometer},\ }\href
	{https://doi.org/10.1093/nsr/nwaa194} {\bibfield  {journal} {\bibinfo
			{journal} {Nat. Sci. Rev.}\ }\textbf {\bibinfo {volume} {8}},\ \bibinfo
		{pages} {nwaa194} (\bibinfo {year} {2020}{\natexlab{b}})}\BibitemShut
	{NoStop}%
	\bibitem [{\citenamefont {Liu}\ and\ \citenamefont
		{Wang}(2022)}]{liu2022arxiv}%
	\BibitemOpen
	\bibfield  {author} {\bibinfo {author} {\bibfnamefont {Y.}~\bibnamefont
			{Liu}}\ and\ \bibinfo {author} {\bibfnamefont {Z.-Y.}\ \bibnamefont {Wang}},\
	}\bibfield  {title} {\bibinfo {title} {{Shortcuts to Adiabaticity with
				Inherent Robustness and without Auxiliary Control}},\ }\href
	{https://arxiv.org/abs/2211.02543} {\bibfield  {journal} {\bibinfo  {journal}
			{arXiv:2211.02543}\ } (\bibinfo {year} {2022})}\BibitemShut {NoStop}%
	\bibitem [{\citenamefont {Carollo}\ \emph {et~al.}(2006)\citenamefont
		{Carollo}, \citenamefont {Santos},\ and\ \citenamefont
		{Vedral}}]{Carollo2006}%
	\BibitemOpen
	\bibfield  {author} {\bibinfo {author} {\bibfnamefont {A.}~\bibnamefont
			{Carollo}}, \bibinfo {author} {\bibfnamefont {M.~F.}\ \bibnamefont
			{Santos}},\ and\ \bibinfo {author} {\bibfnamefont {V.}~\bibnamefont
			{Vedral}},\ }\bibfield  {title} {\bibinfo {title} {{Coherent Quantum
				Evolution via Reservoir Driven Holonomies}},\ }\href
	{https://doi.org/10.1103/PhysRevLett.96.020403} {\bibfield  {journal}
		{\bibinfo  {journal} {Phys. Rev. Lett.}\ }\textbf {\bibinfo {volume} {96}},\
		\bibinfo {pages} {020403} (\bibinfo {year} {2006})}\BibitemShut {NoStop}%
	\bibitem [{\citenamefont {Rivas}\ and\ \citenamefont {Huelga}(2012)}]{rivas}%
	\BibitemOpen
	\bibfield  {author} {\bibinfo {author} {\bibfnamefont {A.}~\bibnamefont
			{Rivas}}\ and\ \bibinfo {author} {\bibfnamefont {S.~F.}\ \bibnamefont
			{Huelga}},\ }\href@noop {} {\emph {\bibinfo {title} {Open Quantum Systems}}}\
	(\bibinfo  {publisher} {Springer},\ \bibinfo {address} {Berlin},\ \bibinfo
	{year} {2012})\BibitemShut {NoStop}%
	\bibitem [{SM()}]{SM}%
	\BibitemOpen
	\href@noop {} {}\bibinfo {note} {See Supplemental Material for more
		details.}\BibitemShut {Stop}%
	\bibitem [{\citenamefont {Dobrovitski}\ \emph {et~al.}(2009)\citenamefont
		{Dobrovitski}, \citenamefont {Feiguin}, \citenamefont {Hanson},\ and\
		\citenamefont {Awschalom}}]{dobrovitski2009spinnoise}%
	\BibitemOpen
	\bibfield  {author} {\bibinfo {author} {\bibfnamefont {V.~V.}\ \bibnamefont
			{Dobrovitski}}, \bibinfo {author} {\bibfnamefont {A.~E.}\ \bibnamefont
			{Feiguin}}, \bibinfo {author} {\bibfnamefont {R.}~\bibnamefont {Hanson}},\
		and\ \bibinfo {author} {\bibfnamefont {D.~D.}\ \bibnamefont {Awschalom}},\
	}\bibfield  {title} {\bibinfo {title} {{Decay of Rabi Oscillations by
				Dipolar-Coupled Dynamical Spin Environments}},\ }\href
	{https://doi.org/10.1103/PhysRevLett.102.237601} {\bibfield  {journal}
		{\bibinfo  {journal} {Phys. Rev. Lett.}\ }\textbf {\bibinfo {volume} {102}},\
		\bibinfo {pages} {237601} (\bibinfo {year} {2009})}\BibitemShut {NoStop}%
\end{thebibliography}

%

\end{document}